\documentclass[10pt,twocolumn]{article}

\usepackage{graphicx}
\usepackage{indentfirst,csquotes}
\usepackage{amssymb,amsthm,amsmath}
\usepackage{xcolor,paralist,hyperref,titlesec,fancyhdr}
\usepackage{lipsum}
\usepackage{siunitx}    
\usepackage[numbers,sort&compress]{natbib}
\usepackage{booktabs}   
\usepackage{float}
\usepackage{braket} 
\usepackage{multirow}

\usepackage[
  letterpaper,
  left=0.75in,right=0.75in,
  top=0.80in,bottom=0.80in,
  headheight=13.6pt,
  headsep=2pt,
  footskip=14pt
]{geometry}

\titleformat{\section}{\normalfont\large\bfseries}{\thesection}{1em}{}
\titlespacing*{\section}{0pt}{1ex}{1ex}

\hypersetup{ colorlinks=true, linkcolor=black, filecolor=black, urlcolor=black, citecolor=blue }

\titleformat{\section}
  {\centering\normalfont\normalsize\bfseries}
  {\thesection.}
  {0.6em}
  {\MakeUppercase}
\titlespacing*{\section}{0pt}{3.0ex}{0.8ex}

\titleformat{\subsection}
  {\centering\normalfont\normalsize\bfseries}
  {\thesubsection.}
  {0.6em}
  {}
\titlespacing*{\subsection}{0pt}{2.5ex}{0.8ex}

\makeatletter
\newcommand{\captionsize}{9.0}
\newcommand{\captionbaselineskip}{10.0}

\renewcommand\fnum@figure{Figure~\thefigure.}
\renewcommand\fnum@table {Table~\thetable.}

\long\def\@makecaption#1#2{%
  \vskip\abovecaptionskip
  \begingroup
  \rmfamily\fontsize{\captionsize}{\captionbaselineskip}\selectfont
  \sbox\@tempboxa{#1~#2}%
  \ifdim\wd\@tempboxa>\hsize
    \noindent #1~#2\par
  \else
    \hb@xt@\hsize{\hfil\box\@tempboxa\hfil}%
  \fi
  \endgroup
  \vskip\belowcaptionskip
}
\makeatother

\makeatletter
\let\oldtable\table
\let\endoldtable\endtable
\renewenvironment{table}[1][tbp]{
  \oldtable[#1]
  \footnotesize 
}{\endoldtable}
\makeatother

\begin{document}

\twocolumn[
\begin{@twocolumnfalse}
\begin{center}
{\large\bfseries Phase Dynamics of Self-Accelerating Bose--Einstein Condensates\par}
\vspace{0.8ex}
{\normalsize Maximilian L. D. D. Pellner$^{1}$ and Georgi Gary Rozenman$^{2}$\par}

\vspace{0.4ex}
{\small
$^{1}$Fakultät für Physik, Ludwig-Maximilians-Universität, München, Germany\\
$^{2}$Department of Mathematics, Massachusetts Institute of Technology, Cambridge, Massachusetts 02139, USA\\

\today
\par}
\end{center}

\vspace{-2.8ex}
\begin{center}
\begin{minipage}{0.80\textwidth}
\small\hspace*{0.3cm} Self-accelerating Airy matter waves offer a clean setting to access the intrinsic cubic-in-time phase. Here we reconstruct the relative phase of simulated Airy-shaped Bose–Einstein condensates from interference fringes in free space, a regime approached in microgravity. The cubic phase dynamics are quantified via approximately debiased, windowed polynomial fits with systematics-aware uncertainty estimates that account for window-induced correlations.
We compare two physically feasible phase-extraction methods – heterodyne-based and density-based – and show that an Airy-Gaussian geometry yields substantially improved robustness to fit-window selection relative to an Airy-Airy collision.
In the weakly interacting regime, the extracted cubic coefficient responds linearly to leading order to the effective interaction strength, its shift from the noninteracting value providing a calibrated probe of weak mean-field nonlinearities in self-accelerating condensates.
\end{minipage}
\end{center}
\vspace{1.2ex}

\end{@twocolumnfalse}
]

\thispagestyle{empty}

\section{Introduction}

Since their first experimental realizations in dilute atomic gases in 1995 \cite{ketterle1995,cornell1995},
Bose-Einstein condensates (BECs) have emerged as a versatile platform for 
studying macroscopic quantum phenomena~\cite{dalfovo1999,leggett2001,pethick2002,pitaevskii2003}. 
At ultracold temperatures, a large fraction of the atoms occupy a single 
quantum state, giving rise to coherence properties analogous to those of 
lasers in optics \cite{andrews1997,mewes1997}. This unique feature has made BECs a fertile ground for 
exploring fundamental questions in quantum mechanics, nonlinear wave dynamics \cite{carretero2008}, 
and analog models of condensed matter \cite{bloch2008} and cosmology \cite{barcelo2011}.
At the theoretical level, their mean-field dynamics are widely captured by the Gross--Pitaevskii equation, with driven--dissipative generalizations extending this framework to other quantum fluids such as exciton--polariton and photonic condensates \cite{deng2010,carusotto2013}.

The vast majority of studies to date has focused on condensates prepared in ground states of harmonic traps, which are often well approximated by Gaussian-like profiles in the weakly interacting limit \cite{dalfovo1999,pethick2002}.
Such simple ground states provide a natural starting point for investigating equilibrium properties, collective excitations, and nonlinear interactions. 
Upon release from the trap, the ensuing time-of-flight expansion of such states is well understood in terms of ballistic, essentially self-similar expansion, providing a standard baseline for interferometric and dynamical studies \cite{castin1996,kagan1996,andrews1997}.
However, the restriction to near-Gaussian ground states limits the exploration of more exotic dynamics that can arise from nontrivial initial conditions \cite{dalfovo1999,pethick2002}.

In parallel, structured wave packets beyond near-equilibrium ground states have attracted long-standing interest across quantum mechanics and wave physics.
A particularly striking example is the Airy wave packet examined by M. V. Berry and N. L. Balazs, 
which is characterized by its unique dispersionless evolution maintaining its functional form, while its caustic follows a parabolic trajectory even in the absence of an external~force~\cite{berrybalasz1979}.

\begin{figure}[h]
  \centering
  \includegraphics[width=\columnwidth, trim=100 1 222 102,clip
  ]{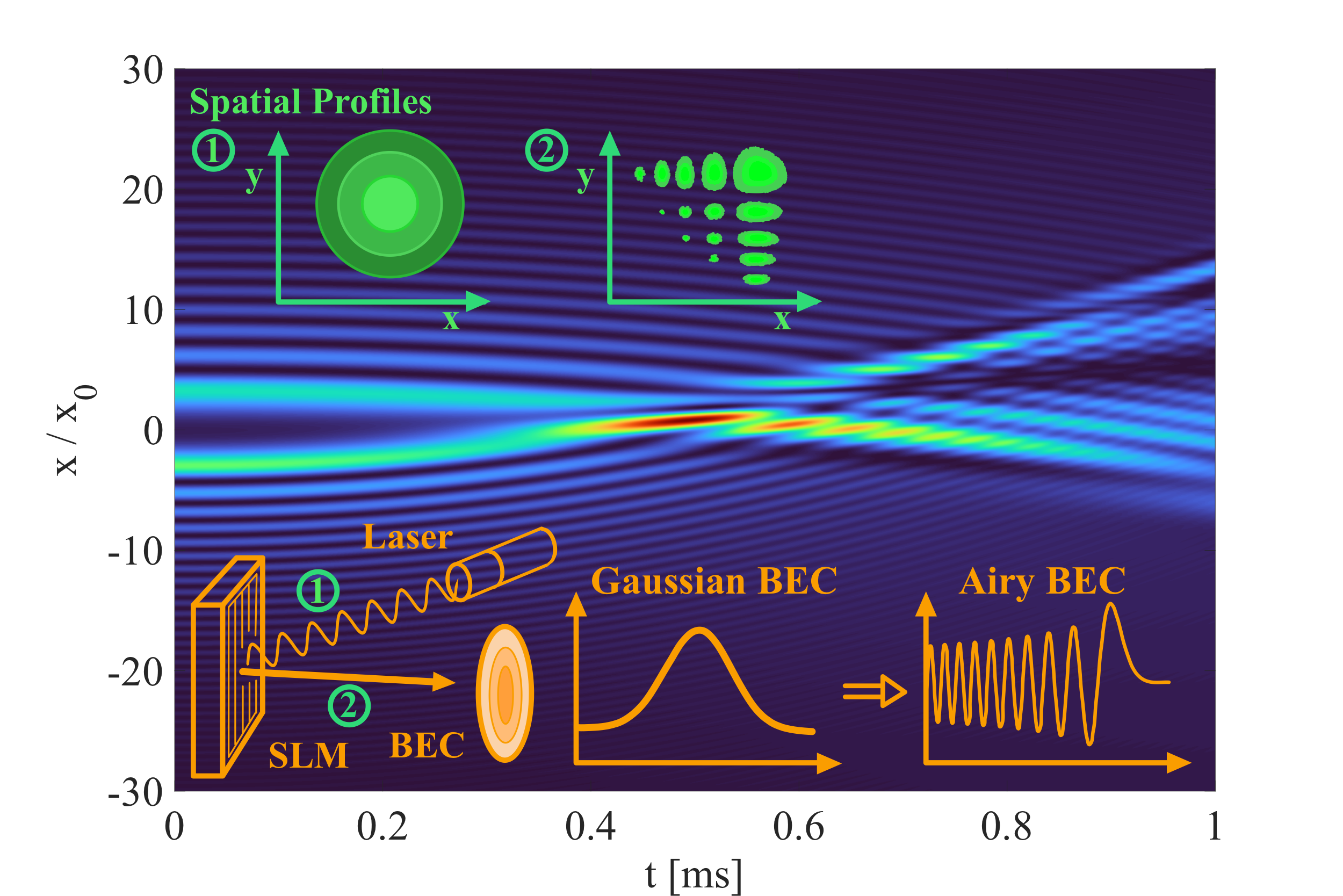}                  
  \caption{Time evolution of density $|\Psi(x,t)|^2$: Interference of two counterpropagating Airy-Bose-Einstein Condensates (Ai-BECs) in free space. A slightly asymmetric scenario of the Airy scales $x_{0,1}$, $x_{0,2}=1.3 \cdot x_{0,1}$ is chosen, otherwise the cubic phase dynamics cancel out. \newline
  \emph{Inset:} schematic of the preparation protocol (Appendix~\ref{app:preparation}): a spatial light modulator and a Fourier lens shape a laser mode into a finite-energy Airy beam, whose amplitude is written onto the Gaussian BEC ground state by a weak two-photon Raman pulse, producing an Ai-BEC~\cite{efremidis2013}. Sign-mirrored masks produce two packets that self-accelerate toward each other, as shown, and a final state-mixing pulse restores the interference for readout. The two-dimensional beam is drawn for illustration only, the analysis being effectively one-dimensional along~$x$.}
  \label{fig:airy_collision}
\end{figure}                          

Because the ideal Airy solution carries infinite energy, physical realizations rely on finite-energy truncations. In optics these were introduced and experimentally observed as accelerating Airy beams \cite{siviloglou2007ol,siviloglou2007prl}, which also exhibit self-healing after partial obstruction \cite{broky2008}.
Airy-type wave packets have since been demonstrated in other diffractive and dispersive wave systems, including free-electron beams \cite{volochbloch2013}, with the broader field of accelerating waves reviewed in Ref.~\cite{efremidis2019overview}.

Complementarily, wave packets of arbitrary shape evolving in a linear potential acquire a spatially uniform phase term that is cubic in time. 
This global cubic phase factor appears in the quantum-mechanical propagator of a particle subject to a constant force and was already derived by E.~H.~Kennard in 1927 \cite{kennard1927,zimmermann2017}. 
This force-induced, spatially uniform cubic-in-time phase is characteristic of any packet subject to a constant force \cite{zimmermann2017,rozenman2019amplitude}.
Airy wave packets also inherently exhibit cubic phase dynamics independent of the position, even in the absence of an external potential as already apparent in the Berry--Balazs solution \cite{berrybalasz1979}.
This deep intrinsic connection between this property and the former quantum mechanical evolution~\cite{pellner2026avs,pellner2026thesis} is consistent with the fact that Airy functions are eigenstates of such linear potentials \cite{landau_nonrelativistic}.
Three cubic-in-time phases are kept distinct throughout: the force-induced phase of an arbitrary packet in a linear potential \cite{kennard1927}, the intrinsic phase of a freely propagating Airy packet in the absence of any force \cite{berrybalasz1979}, which is the object studied here, and the relative cubic phase between two packets that alone is accessible in an interference measurement.
This force-induced cubic phase has been observed in surface-gravity water waves, which provide a quantum-mechanical analog platform governed by a Schrödinger-type equation with an effective linear potential \cite{rozenman2019amplitude,rozenman2019quantum,Rozenman2021Projectile,rozenman2020observation}.

The same cubic-in-time phase term underpins $T^3$ matter-wave interferometry schemes based on state-dependent forces \cite{zimmermann2017}.
Recently, this force-induced cubic phase was experimentally realized in an atom-chip-based Stern--Gerlach matter-wave interferometer \cite{amit2019}.
In microgravity, a near-free-fall environment with well-characterized residual gradients provides a clean setting in which the following free-space interferometric collisions can, in principle, be implemented \cite{Raudonis2023MicrogravityFacilities,Williams2024CALPathfinder}.

These developments motivate exploring Airy-shaped initial states in ultracold quantum gases~\cite{yuce2015selfaccel,koehl2001}, where mean-field interactions provide a controlled nonlinearity~\cite{qin2019acceleratebec} and interference readout offers access to the evolving relative phase.
However, standard cold-atom imaging measures the density, and phase – especially spatially uniform terms – can only be accessed as a relative phase in an interferometer \cite{andrews1997,zhai2018talbot}, making it sensitive to global and spatially varying phase drifts and to density-dependent mean-field phase accumulation \cite{dalfovo1999,pethick2002,Cronin2009}. 
Using the noninteracting evolution as a benchmark, the following simulations show that the cubic-in-time phase responds sensitively to the mean-field interaction: its coefficient varies linearly with interaction strength in the perturbative regime, as expected at leading order, and crosses over to a distinct nonlinear regime at stronger interactions.

The guiding line of this work is a single, practically motivated goal: to extract the intrinsic cubic-in-time phase of self-accelerating Airy condensates robustly and in a nearly unbiased manner from realistic interference data, and to turn it into a controlled probe of weak mean-field nonlinearity.
The interference of Airy-shaped BEC wave packets (Ai-BECs), illustrated in Fig.~\ref{fig:airy_collision}, is analyzed to predict the intrinsic cubic phase dynamics in the weakly interacting regime.
It is shown that interferometric data can be demodulated to isolate the intrinsic, spatially uniform cubic-in-time phase and access it as a relative phase in an interacting condensate.
Based on simulations, two phase-reconstruction procedures – heterodyne-based and density-based – are proposed and benchmarked.
The sensitivity to the choice of analysis window, and the resulting bias, are quantified by scanning the fit interval.
To construct a nearly unbiased phase estimator in the perturbative regime, the procedure is calibrated via the analytic linear case $g=0$, and uncertainties are estimated using heteroskedasticity and autocorrelation consistent Newey–West standard errors.
Thus, the study provides theoretical guidance and simulation-based motivation for extending ultracold-gas interferometry beyond near-Gaussian initial 
states to self-accelerating, nearly non-spreading and self-healing quantum-fluid matter waves over finite times \cite{berrybalasz1979,siviloglou2007prl,broky2008,malomed2022selfaccel}.

\section{Theoretical Model}

The dynamics of a dilute Bose-Einstein condensate at ultracold temperatures 
is accurately described by the Gross--Pitaevskii equation (GPE), a nonlinear 
Schrödinger equation that incorporates mean-field interactions between the atoms \cite{dalfovo1999,pethick2002,pitaevskii2003}, 
\begin{equation}
\begin{aligned}
    \mathrm{i}\hbar \frac{\partial \Psi(\mathbf{r},t)}{\partial t} = \left[ -\frac{\hbar^2}{2m} \nabla^2 + V(\mathbf{r},t) + g_{\mathrm{3D}} |\Psi(\mathbf{r},t)|^2 \right] \\[0.5ex]
     \cdot \Psi(\mathbf{r},t) \,, 
\end{aligned}
\end{equation}
where \(\Psi(\mathbf{r},t)\) is the condensate wavefunction normalized such that 
\(\int |\Psi|^2 d\mathbf{r} = N\), with \(N\) the number of atoms. Here, 
\(m\) is the atomic mass, \(V(\mathbf{r},t)\) is an external potential, 
and $g_{3\mathrm{D}} = 4\pi \hbar^2 a_\mathrm{s}/m$ is the three-dimensional interaction strength with \(a_\mathrm{s}\) the s-wave scattering length \cite{dalfovo1999,pethick2002}.

In the weakly interacting regime, the dynamics are well captured at leading order by the linear Schrödinger equation, providing an analytic reference \cite{dalfovo1999,pethick2002}. 
In the simulations, the full time evolution is obtained with a split-step method. The residual interaction effects are included via the nonlinear phase shift 
$\Psi(x,t) \mapsto \exp[-\mathrm{i}(g/\hbar)|\Psi(x,t)|^2\,\Delta t]\,\Psi(x,t)$
with the normalization $\int |\Psi(x,t)|^2\,\mathrm{d}x = 1$ \cite{javanainen2006}.
Because the wavefunction is normalized to unity rather than to the atom number $N$, this effective coupling absorbs the atom number, $g \equiv g_\mathrm{eff} = N\,g_{1\mathrm{D}}$, with $g_{1\mathrm{D}}$ the per-atom one-dimensional coupling. Unless stated otherwise, $g$ denotes this effective one-dimensional control parameter used in the numerics, and with this normalization $g/\hbar$ carries units of $\mathrm{m/s}$.
In what follows, for convenience, a quasi-one-dimensional condensate is considered \cite{olshanii1998}, where the dynamics are effectively 
restricted to the \(x\) direction,
\begin{equation}
    \mathrm{i}\hbar\,\frac{\partial}{\partial t}\Psi(x,t)
  =
  \left[
    -\frac{\hbar^{2}}{2m}\,\frac{\partial^{2}}{\partial x^{2}}
    + V(x)
  \right] \Psi(x,t).
\end{equation}

By choosing a linear potential $V = Fx$, 
and using the ansatz $\Psi(x,t) = \psi(x) \; e^{-\mathrm{i} \mu t / \hbar }$ of the stationary eigenstates, the linear Schrödinger equation simplifies to the following equation with $l = \frac{\mu}{F}$:
\begin{equation}
    \frac{\mathrm{d}^{2}\psi}{\mathrm{d}x^{2}}
    - \frac{2mF}{\hbar^{2}}\,
    \bigl(x - l\bigr)\,\psi(x)
    = 0.
    \end{equation}
Introducing the characteristic length scale $x_0 = \left( \frac{\hbar^2}{2mF} \right)^{1/3} $ and the dimensionless variables $\xi =\frac{x-l}{x_0}$ and $\tau = \frac{\hbar \,t}{m \, x_0^2}$ for the subsequent time propagation, one obtains the Airy equation \cite{landau_nonrelativistic}:
\begin{equation}
    \frac{\mathrm{d}^2\psi}{\mathrm{d}\xi^2} - \xi \psi(\xi) = 0 \;.
\end{equation}

The corresponding Airy eigenstates are $\operatorname{Ai}(\xi)$ and $\operatorname{Bi}(\xi)$, of which the former is physically admissible, since $\operatorname{Bi}(\xi)$ diverges for $\xi \to +\infty$ and is therefore not square-integrable \cite{landau_nonrelativistic,abramowitz1964handbook}. 
The profound Fourier representation \cite{berrybalasz1979} is particularly relevant in the present context, as it reveals the rich structure of the nontrivial momentum distribution,
\begin{equation}
  \operatorname{Ai}(\xi)
  = \frac{1}{2\pi}
    \int_{-\infty}^{\infty}
    \exp\!\left[
      \mathrm{i}\left( \frac{k^{3}}{3} + \xi k \right)
    \right] \,\mathrm{d}k \, .
\end{equation}

\subsection{Single Airy Propagation and Phase Dynamics}
The linear potential introduced above serves only to select the Airy eigenstate and fix its length scale $x_0$, or equivalently the eigenforce that drives the free self-acceleration and parametrizes the cubic-phase landscape in a companion study~\cite{pellner2026avs,pellner2026thesis}. The dynamics studied here are those of free-space propagation, for which the accumulated cubic-in-time phase is intrinsic. Equivalently, in the frame comoving with the caustic the free packet is the stationary Airy eigenstate, the selecting linear potential reappearing as a constant fictitious force~\cite{berrybalasz1979}.
In order to analytically derive the phase dynamics for $g=0$, the Feynman propagator convolution is computed.
The wave packet is evolved by evaluating
\begin{equation}
    \Psi(\xi,\tau) = \int \mathrm{d}\xi' \ K(\xi,\tau,\xi',0) \ \psi(\xi')            
\end{equation}
as the integral of the wave packet itself multiplied by the Feynman propagator of free propagation
\begin{equation}
    K(\xi,\tau;\xi',0)= \frac{1}{\sqrt{2\pi \mathrm{i} \tau}} \                 
    \exp\!\left[\frac{\mathrm{i} (\xi - \xi')^2}{2\tau}\right]
\end{equation}
\cite{schleich2011quantum, feynman1965qm_path-integrals}.
For the experimental realization, the practical model 
is the exponentially truncated first Airy solution, which has finite energy \cite{siviloglou2007ol} and therefore is physically meaningful and numerically computable,
\begin{equation}
    \psi_{\mathrm{Ai}}(\xi) = C \ \mathrm{Ai}(\xi) \ e^{a\xi} \, , \ \ a >0.
\end{equation}
After a few manipulations of the integral - in particular completing the square in the exponent, evaluating the Gaussian integral, performing a change of variables for a transverse shift and rewriting the result, Eq. (6) yields

\begin{equation}
\begin{aligned}
\Psi_{\mathrm{Ai}}(\xi,\tau)
  &= C \exp\!\Bigl[
        - \frac{a\tau^2}{2}
        + a\xi
        + \mathrm{i}\Bigl(
            -\frac{\tau^3}{12}
            + \frac{\tau\xi}{2}
        \Bigr.\\[-0.2ex]
  &\qquad\Bigl.
            + \frac{a^2\tau}{2}
        \Bigr)
     \Bigr]
     \cdot \mathrm{Ai}\!\left(\xi - \frac{\tau^2}{4} + i a \tau\right) .
\end{aligned}
\end{equation}

Finally, the relevant phase – including the intrinsic cubic-in-time phase, the characteristic cubic term in
$\tau$ – can be extracted straightforwardly,
\begin{equation}
    \Phi_{Ai}(\xi,\tau) = -\frac{\tau^3}{12} + \frac{\tau\xi}{2} + \frac{a^2\tau}{2},
\end{equation}
\cite{siviloglou2007ol}, where the imaginary component in the argument of the Airy function is neglected.
Intuitively, for sufficiently large numerical simulation domains, the apodization parameter
$a$ can be chosen small enough that its influence on the phase dynamics becomes negligible. 
In the limit $a \to 0$, the contribution of the complex shift of the Airy argument vanishes entirely.
For a rigorous discussion of the residual impact at small finite choices of $a$ along the main lobe,
see the supplementary analysis of Rozenman et al \cite{rozenman2019amplitude}.

\subsection{Single Gaussian Propagation and Phase Dynamics}
Analogously to the previous calculation, the evolution of a stationary, i.e. zero-momentum, Gaussian wave packet 
\begin{equation}
    \Psi_{\mathrm{G,0}}(\xi)= \frac{1}{\left( 2 \pi s^2 \right)^{\frac{1}{4}}} \ \exp\left[ -\frac{\xi^2}{4s^2} \right]
\end{equation}
is obtained via convolution with the Feynman propagator, 
where $s = \frac{\sigma_0}{x_0}$ 
denotes the dimensionless normalized width at $t=0$, and
$\eta = \frac{\tau}{2s^2}$.
Thereby, the resulting time-dependent general solution reads
\begin{equation}
\begin{aligned}
\Psi_{\mathrm{G}}(\xi,\tau)
  &= \frac{1}{\left( 2 \pi \left( s \sqrt{1+\mathrm{i} \eta} \right)^2 \right)^{\frac{1}{4}}} \\[0.5ex]
  &\quad \cdot \exp\!\left[ -\frac{\xi^2}{4 \left( s \sqrt{1+\mathrm{i}\eta}\right)^2} \right].
\end{aligned}
\end{equation}
The effective time-dependent normalized width is given by $s_{\mathrm{c}}(\tau)= s \, \sqrt{1+\mathrm{i}\eta}$.
Furthermore, Eq. 12 can be transformed into 
\begin{equation}
\begin{aligned}
    \Psi_{\mathrm{G}}(\xi,\tau)
    &= \frac{1}{\left( 2 \pi \left( s \sqrt{1+\eta^2} \right)^2 \right)^{\frac{1}{4}}} \\[0.5ex]
    &\quad \cdot \exp\!\left[ \mathrm{i}\left( \frac{\eta\xi^2}{4 s^2 (1+\eta^2)} -\frac{1}{2} \arctan(\eta) \right) \right] \\[0.5ex]
    &\quad \cdot \exp\!\left[ -\frac{\xi^2}{4 s^2 \left(1+\eta^2\right)} \right],
\end{aligned}
\end{equation}
where $s_{\tau}(\tau)= s \, \sqrt{1+ \eta^2}$ 
is the corresponding time-dependent width \footnote{Noting that the previously defined width $s_{\mathrm{c}}(\tau)$ is in general a complex quantity, hence this second definition is real.}.
After collecting the imaginary contributions from the prefactor and the complex exponent – the former is referred to as the Gouy phase,
the phase dynamics are described by 
\begin{equation}
    \Phi_\mathrm{G}(\xi,\tau)= \frac{\eta \xi^2}{4s^2 (1+\eta^2)} - \frac{1}{2} \arctan(\eta).                  
\end{equation}
For early times, when $\tau$ is sufficiently small,  
the Taylor expansion $\arctan(\eta)= \eta - \frac{1}{3}\eta^3 + \mathcal{O}(\eta^5)$ 
provides more interpretable insight into the specific polynomial phase behavior, yielding
\begin{equation}
    \Phi_\mathrm{G}(\xi,\tau)= \frac{\xi^2 \tau}{2 (4s^4+\tau^2)} - \frac{\tau}{4s^2} + \frac{\tau^3}{48s^6}.
\end{equation}

\subsection{Airy--Airy Collision}
To circumvent the limitation that only the density distribution of single quantum systems can be directly accessed in quantum mechanics,
the interference of two wave packets is investigated.
By carefully analyzing the structure of the interference fringes, information about the underlying phase dynamics can be extracted.
One simple interference scenario is the collision of two counterpropagating Ai-BECs in free space,
since they naturally accelerate towards each other without requiring any initial momentum or external potential.

The superposition of two independent quantum mechanical objects, here chosen $\Psi_1=\Psi_{Ai,1}$, $\Psi_2=\Psi_{Ai,2}$ w.l.o.g., is described by
\begin{equation}
    \Psi_{tot} = \Psi_{\mathrm{Ai,1}} + \Psi_{\mathrm{Ai,2}}.
\end{equation}
Thus, the measurement of the probability density is 
\begin{equation}
\begin{aligned}
    |\Psi_{\mathrm{tot}}|^2=|\Psi_{\mathrm{Ai,1}}|^2 +|\Psi_{\mathrm{Ai,2}}|^2 + 2|\Psi_{\mathrm{Ai,1}}||\Psi_{\mathrm{Ai,2}}| \\[0.5ex]
    \cdot \cos\left(\Delta\Phi(\xi,\tau)\right),
\end{aligned}
\end{equation}
with relative phase  $\Delta\Phi(\xi,\tau)= \arg(\Psi_{\mathrm{Ai,1}}) - \arg(\Psi_{\mathrm{Ai,2}})$.
For this particular experiment, the phase dynamics are
\begin{equation}
\begin{aligned}
    \Delta\Phi_{\mathrm{Ai-Ai}}(\xi,t)= - \frac{\tau_1^3}{12} + \frac{\tau_1 \,\xi_1}{2} + \frac{a_1^2\tau_1}{2} \\[0.5ex]
    + \frac{\tau_2^3}{12} -\frac{\tau_2 \,\xi_2}{2} - \frac{a_2^2\tau_2}{2} \\[0.5ex]
\end{aligned}
\end{equation}
with $\tau_i = \frac{t}{\tau_{0,i}}$, where $\tau_{0,i}$ ($i=1,2$) sets the individual Airy scale of each packet.
Transforming the characteristic time scales according to $\tau_{0,i} = m x_{0,i}^2/\hbar$, the leading cubic coefficient reads
\begin{equation}
    c_{\mathrm{3, Ai-Ai}} = \frac{\hbar^3}{12m^3} \left( \frac{1}{x_{0,2}^6}-\frac{1}{x_{0,1}^6}\right).
\end{equation}

\subsection{Airy--Gaussian Collision}
Another configuration that appears within experimental reach is the collision of an Airy and a Gaussian wave packet,
again without requiring any initial momentum or external potential.
The resulting phase difference is
\begin{equation}
\begin{aligned}
    \Delta\Phi_{\mathrm{Ai-G}}(\xi,t)
    &= + \frac{\tau_{\mathrm{Ai}}^3}{12} +\frac{\tau_{\mathrm{G}}^3}{48 s^6}
       - \frac{\tau_\mathrm{G}}{4s^2} - \frac{\tau_{\mathrm{Ai}}\xi_{\mathrm{Ai}}}{2} \\[0.5ex]
    &\quad - \frac{a^2\tau_{\mathrm{Ai}}}{2}
       + \frac{\xi_{\mathrm{G}}^2 \tau_\mathrm{G}}{2 (4s^4+\tau_{\mathrm{G}}^2)} \, 
\end{aligned}
\end{equation}
with $\tau_{\mathrm{Ai}} = \frac{t}{\tau_{\mathrm{0,Ai}}}$ and $\tau_\mathrm{G} = \frac{t}{\tau_{\mathrm{0,G}}}$.
Accordingly, upon restoring physical units, the spatially uniform cubic coefficient is given by
\begin{equation}
    c_{\mathrm{3, Ai-G}} = \frac{\hbar^3}{m^3x_0^6} \left(\frac{1}{12} +\frac{1}{48s^6}\right).
\end{equation}

\subsection{Heterodyne Phase Extraction}
Extracting the phase could theoretically be performed directly via accessing the phase of the separately propagated wave packet amplitudes
in the simulation, which are later added for the superposition. 
However, a physically feasible way to obtain the phase dynamics is used here.
In the simulation, $\Psi_2^{*}\Psi_1$ is available directly from the propagated amplitudes and serves as the ground-truth reference against which the reconstructed signal below is validated.
Motivated by the interference contribution $2\,\Re[\Psi_2^{*}\,\Psi_1]$ to the density of two overlapping condensates \cite{pethick2002},
the complex signal $S_\mathrm{HPE}$ is defined as a windowed overlap integral that characterizes the coherence between the two wave packets,
\begin{equation}
    S_{\mathrm{HPE}}(\tau) = \int w(\xi) \; \Psi_2^*(\xi,\tau) \; \Psi_1(\xi,\tau) \; e^{-\mathrm{i}k_\mathrm{f}(\tau)(\xi-\xi_\mathrm{c}(\tau))} \; \mathrm{d}\xi,
\end{equation}
where $w(\xi) = \exp\left( -\frac{1}{2} \frac{(\xi - \xi_\mathrm{c}(\tau))^2}{\sigma_{\mathrm{ROI}}^2} \right)$ denotes a Gaussian envelope as the region of interest (ROI) centered on the current overlap peak $\xi_\mathrm{c}(\tau)$, hence implicitly time-dependent, with a width $\sigma_{\mathrm{ROI}}$ of order $10^{-2}$ to $10^{-1}$ in dimensionless units,
and $k_\mathrm{f}(\tau) = k_{\mathrm{c,1}}(\tau) - k_{\mathrm{c,2}}(\tau)$ as the differential fringe carrier, time-dependent and originating from a local plane-wave approximation discussed in the next subsection.
This approach extends simple overlap measurements by incorporating a Gaussian spatial filter $w(\xi)$ and Fourier-based fringe-carrier demodulation~\cite{Takeda}, here in a centered, symmetrized form introduced for robustness to carrier misdetermination, enabling localized sensitive phase extraction. The spatial integration removes only the fringe carrier $k_\mathrm{f}$, whereas the spatially uniform cubic-in-time phase, being independent of $\xi$, factors out of the integral and is retained in $\arg S_\mathrm{HPE}$. Removing the carrier maximizes $|S_\mathrm{HPE}|$, so that its argument, and with it the cubic-in-time phase, stays well defined and numerically stable to extract.

On the level of the wave function, the amplitudes entering $S_\mathrm{HPE}$ are not directly accessible in an experiment, where only the density is measured.
The complex cross term $\Psi_2^{*}\Psi_1$ is reconstructed by phase-shifting interferometry: a fixed set of density images with controlled, spatially uniform relative-phase offsets $\alpha$ on one packet yields the in-phase and quadrature components of the fringe, and thus $\Psi_2^{*}\Psi_1$ in full, a standard fixed-phase-step reconstruction technique~\cite{bruning1974,degroot2015}.\footnote{In the standard four-step point-sampling scheme the offsets $\alpha\in\{0,\pi/2,\pi,3\pi/2\}$ yield $\rho_\alpha = |\Psi_1|^2+|\Psi_2|^2 + 2|\Psi_1||\Psi_2|\cos(\Delta\Phi+\alpha)$, so that the background cancels in the differences and $\Psi_2^{*}\Psi_1 \propto (\rho_0-\rho_\pi) + \mathrm{i}\,(\rho_{3\pi/2}-\rho_{\pi/2})$. The uniform offset $\alpha$ can be implemented by a brief off-resonant light-shift pulse on one packet, a form of optical phase imprinting~\cite{dobrek1999}. Denschlag et al.\ demonstrated such selective imprinting on one component of a condensate, the other serving as a phase reference~\cite{denschlag2000}. Since absorption imaging is destructive, each offset requires a separate reproducible run.}
The cost is one separate run per offset, trading the single-shot simplicity of the density-based scheme of the next subsection for direct access to the complex signal.
The desired relative phase is finally obtained as
\begin{equation}
    \phi_{\mathrm{rel}}(\tau) 
    = \mathrm{unwrap}\big(\arg[S_\mathrm{HPE}(\tau)]\big) \, ,
\end{equation}
where $\arg(\cdot)$ denotes the complex argument and $\mathrm{unwrap}(\cdot)$ implements numerical phase unwrapping to enforce temporal continuity.
By construction, the centered carrier demodulation isolates the spatially uniform cubic-in-time phase, separated from the position-dependent fringe terms, as the observable central to this work, a step shared by both the heterodyne signal and the density-based extraction below.

\subsection{Density-based Phase Extraction}

A more practicable approach is the density-based phase extraction, because merely simple density profile measurements are required.
Defining the superposition density
\begin{equation}
    \rho(\xi,\tau)= |\Psi_1|^2 + |\Psi_2|^2 + 2 \, \Re[\Psi_2^* \Psi_1] \, ,
\end{equation}
\cite{pethick2002} and inserting the polar decomposition $\Psi_j(\xi,\tau)= |\Psi_j(\xi,\tau)| \; e^{\,\mathrm{i} \, \Phi_j(\xi,\tau)}$,
one obtains
\begin{equation}
    \rho(\xi,\tau)= |\Psi_1|^2 + |\Psi_2|^2 + 2 \; A(\xi,\tau) \, \cos(\Delta\Phi(\xi,\tau)),
\end{equation}
where $A(\xi,\tau)=|\Psi_1(\xi,\tau)|\,|\Psi_2(\xi,\tau)|$ denotes the local interference amplitude, which is assumed to vary slowly on the scale of the fringe spacing, and $\Delta\Phi(\xi,\tau)=\Phi_1(\xi,\tau)-\Phi_2(\xi,\tau)$ encodes the relative phase dynamics.
To isolate the fringe contribution, a windowed fluctuation signal is defined as
\begin{equation}
    \rho_{\mathrm{fluc}}(\xi,\tau) \approx 
    w(\xi)\Big[\rho(\xi,\tau) - \rho_{\mathrm{mean}}(\tau)\Big] \, ,      
\end{equation}
with the local mean density $\rho_{\mathrm{mean}}(\tau) = \frac{\int w(\xi)\,\rho(\xi,\tau)\,\mathrm{d}\xi} {\int w(\xi)\,\mathrm{d}\xi} \,$ .
In the region of interest, the phase is subsequently locally approximated by a plane wave with a time-dependent offset,
\begin{equation}
    \Delta\Phi(\xi,\tau) \approx 
    k_\mathrm{f}(\tau)\big[\xi-\xi_\mathrm{c}(\tau)\big] + \phi(\tau) \, ,
\end{equation}
which motivates the density-based complex signal
\begin{equation}
    S_{\mathrm{DPE}}(\tau) 
    = \int \rho_{\mathrm{fluc}}(\xi,\tau)\,
      e^{-\mathrm{i}k_\mathrm{f}(\tau)\,[\xi-\xi_\mathrm{c}(\tau)]}\,\mathrm{d}\xi
\end{equation}
Here $S_\mathrm{DPE}$ is the windowed complex Fourier coefficient of the density fluctuation at the carrier $k_\mathrm{f}$, a one-dimensional carrier-demodulation procedure analogous to the Fourier-sideband method of Takeda et al.~\cite{Takeda}.
In addition, it is worth noting that the differential fringe carrier, consisting of $k_{\mathrm{c,1}}$ and $k_{\mathrm{c,2}}$, is obtained from calibration measurements performed on the single wave packets, which allows the extraction of the peak position in momentum space separately at each time $\tau$.\footnote{The extracted carrier is the local wavenumber $k_\mathrm{loc}=\partial_\xi\arg\Psi$. Over the narrow ROI centred at $\xi_\mathrm{c}$ the phase is dominated by its linear part, $\varphi(\xi)\approx\varphi(\xi_\mathrm{c})+\varphi'(\xi_\mathrm{c})(\xi-\xi_\mathrm{c})$, with the curvature $\tfrac12\varphi''(\xi_\mathrm{c})(\xi-\xi_\mathrm{c})^2$ and higher-order terms forming a negligible residual, so $\Psi$ is locally a plane wave of wavenumber $\varphi'(\xi_\mathrm{c})$. In the windowed transform $\mathcal{F}[\Psi](k)=\int w(\xi)\,A(\xi)\,e^{i[\varphi(\xi)-k\xi]}\,\mathrm{d}\xi$ the demodulation $e^{-ik\xi}$ subtracts a linear ramp $k\xi$. At $k=\varphi'(\xi_\mathrm{c})$ the dominant linear part cancels and the integrand reduces to the real envelope $w(\xi)A(\xi)$ carrying only the constant phase $e^{i[\varphi(\xi_\mathrm{c})-\varphi'(\xi_\mathrm{c})\xi_\mathrm{c}]}$, up to the small residual, so its contributions are aligned rather than oscillating. The magnitude cannot exceed the fully aligned sum, $|\mathcal{F}[\Psi](k)|\le\int wA\,\mathrm{d}\xi$, and reaches it only for a constant demodulated phase, hence exactly at $k=\varphi'(\xi_\mathrm{c})$, which defines the carrier $k_{c}$. Any other $k$ retains an uncancelled ramp $(\varphi'(\xi_\mathrm{c})-k)\xi$ that rotates the contributions and falls below the bound. A spatially uniform phase, being independent of $\xi$, factors out as a constant prefactor and leaves this peak, hence $k_{c}$, unaffected.} Each Airy packet's carrier equals its intrinsic main-lobe wavenumber, $k_{\mathrm{c},i}=\tau_i/2$, and hence grows linearly in time, so that $k_\mathrm{f}(\tau)$ is likewise linear in $\tau$. The peak overlap density position $\xi_\mathrm{c}(\tau)$ is visually obvious from the straight, quasi-stationary interference fringe, for example of the two main lobes, and hence this stable and well-defined position can be reliably approximated by the peak position of the total superposition density, owing to the sharp shape of the Airy main lobes – for the Airy-Gaussian collision the procedure is analogous. Eventually, the relative phase is obtained from $S_{\mathrm{DPE}}(\tau)$ via the same argument and phase-unwrapping method as in the previous subsection.

\newpage
\section{Simulation Measurements}

The analytic $g=0$ model provides the exact benchmark: the truncated Airy packet propagates freely and accumulates the intrinsic cubic-in-time phase, whose relative cubic coefficient $c_{\mathrm{3,theo}}$ both extraction methods must recover. The split-step simulation reproduces this result at $g=0$ and, on adding the mean-field nonlinearity, extends it to $g>0$, where the interaction shifts the extracted $c_3$ away from $c_{\mathrm{3,theo}}$, this calibrated shift providing a quantitative probe of the mean-field nonlinearity.
In the following passage, the Ai-Ai and the Ai-G collision phase dynamics are studied
in particular by two physically feasible types of phase extraction, demonstrated here in simulations.

\subsection{Cubic Airy--Airy Phase}
After the phase extraction had been carried out~–~here by selecting a narrow Gaussian ROI with a width of $\sigma_{\mathrm{ROI}} = 0.01$ for the heterodyne phase extraction (HPE) and signal demodulation by removing the linear carrier frequency,
the argument of the resulting complex signal, which is rendered continuous by phase unwrapping, corresponds to the pure relative, global phase.
For quantitative analysis, only connected regions – such as the first interference fringe of the main lobes in Fig.~\ref{fig:airy_collision}, ranging from approximately $0.4  \ \mathrm{ms}$ to $0.6 \ \mathrm{ms}$ \footnote{The initial separation of the two Airy profiles is chosen sufficiently small to increase the temporal overlap of the main-lobe interference fringe. At early times the lobes have accelerated only weakly, so their relative velocity is small, which prolongs the overlap.}
– are considered for modeling the phase dynamics.
The phase data are analyzed using a centered generalized least-squares fit of a third-order polynomial.
This procedure yields the polynomial coefficients, their covariance matrix, and hence their standard errors. \footnote{Only numerical errors enter the analysis here, as no noise model is applied. The reported errors on the fit coefficients therefore constitute a lower bound on realistic experimental uncertainties.}

\begin{figure}[h]
  \centering
  \includegraphics[
    width=\columnwidth,
    trim=25 470 95 32,clip  
  ]{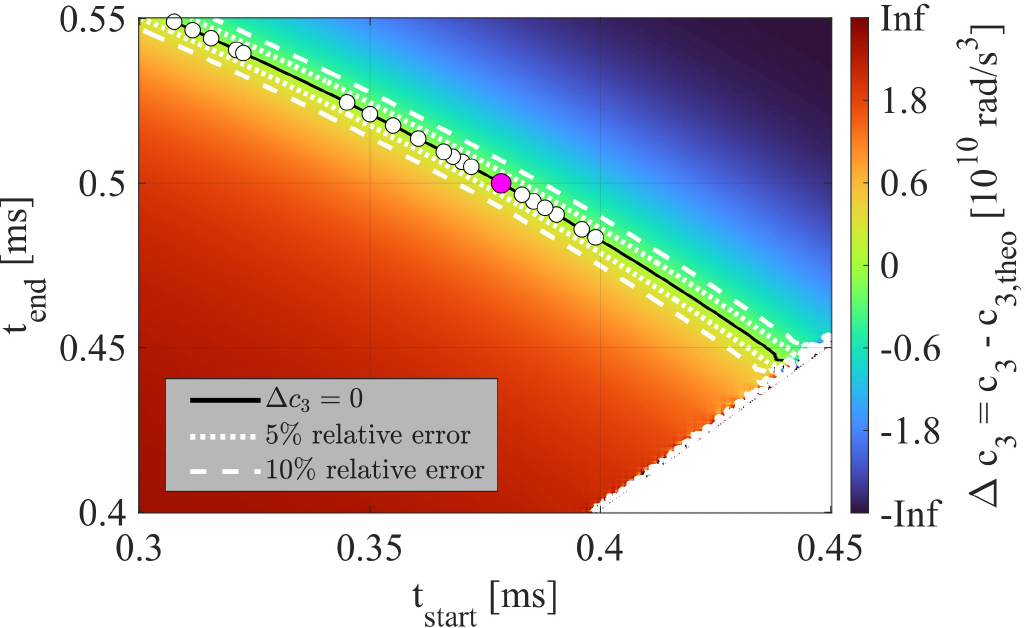}
  \caption{Deviation $\Delta c_3 = c_3 - c_{3,\mathrm{theo}}$ across the fit-window choices $[t_{\mathrm{start}}, t_{\mathrm{end}}]$ for the Ai--Ai collision scenario. 
  The plot visualizes the robustness of the heterodyne phase extraction, i.e. the sensitivity of the cubic-coefficient estimator $c_3$ to the choice of fitted region, which can induce bias. The two white contour lines indicate relative errors of 5\% (dotted) and 10\% (dashed). The highlighted points mark the best discrete window pairs at finite grid resolution. The pink dot denotes the estimate closest to $c_{3,\mathrm{theo}} = -3.472800\times10^{10}\,\mathrm{rad/s^3}$. The colour encodes the signed, nonlinearly compressed deviation $\Delta c_3 = c_3 - c_{\mathrm{3,theo}}$ via
  $T = \operatorname{sgn}\!\big[\tanh(k\,\Delta c_3 / 2s)\big]\ \big|\tanh(k\,\Delta c_3 / 2s)\big|^{\gamma}$,
  with $k = 2.5$, $\gamma = 0.7$, and $s$ given by the 90th percentile of $|\Delta c_3|$, which provides a symmetric, bounded, contrast-enhanced colour scale for $\Delta c_3$ that compresses outliers.}
  \label{fig:ai-ai-heatmap}
\end{figure}

In Fig.~\ref{fig:ai-ai-heatmap}, the deviations of the fitted cubic coefficient $c_3$ from its theoretical value $c_{\mathrm{3,theo}}$, based on the heterodyne simulation phase extraction, are shown. 
The x- and y-axes characterize the choice of the fitting interval, defining each point on the map by one unique pair of the starting and end time of the fitting region. 
The general, global structure is a rather steep valley of minima extending from the upper left corner towards the lower right corner.
The area below this valley exhibits rapidly increasing positive deviations, while the opposite upper region – above the valley – shows increasingly negative deviations, noting that this metaphor focuses on the absolute deviations, disregarding their sign. The contour lines of constant relative deviation of 5 \% and of 10 \% are laid relatively close to the line of optimal fit values.
The top 20 fit values are approximately equally distributed in the upper part, favoring rather larger regions, but this effect might also very likely be caused by the finite numerical resolution.    
The best fitting point is located at $t_{\mathrm{start}}\approx 0.38 \ \mathrm{ms}$, $t_{\mathrm{end}}\approx 0.50 \ \mathrm{ms}$.
Apart from numerical instabilities for very small fitting windows, corresponding to weakly determined fitting conditions along the boundary of the lower, right triangle, the overall landscape is free from random fluctuations or disturbances, providing high confidence in the trend analysis.
The window scan quantifies the sensitivity of the extracted coefficient to the fit interval, thereby characterizing the extraction systematic rather than selecting the interval for agreement with the reference value. The estimator is calibrated once against the analytic $g=0$ coefficient and then applied unchanged at $g\neq0$, so that any residual deviation there reflects genuine nonlinear dynamics.

\begin{figure}[h]
  \centering
  \includegraphics[width=\columnwidth, trim=40 419
  140 88,clip] 
  {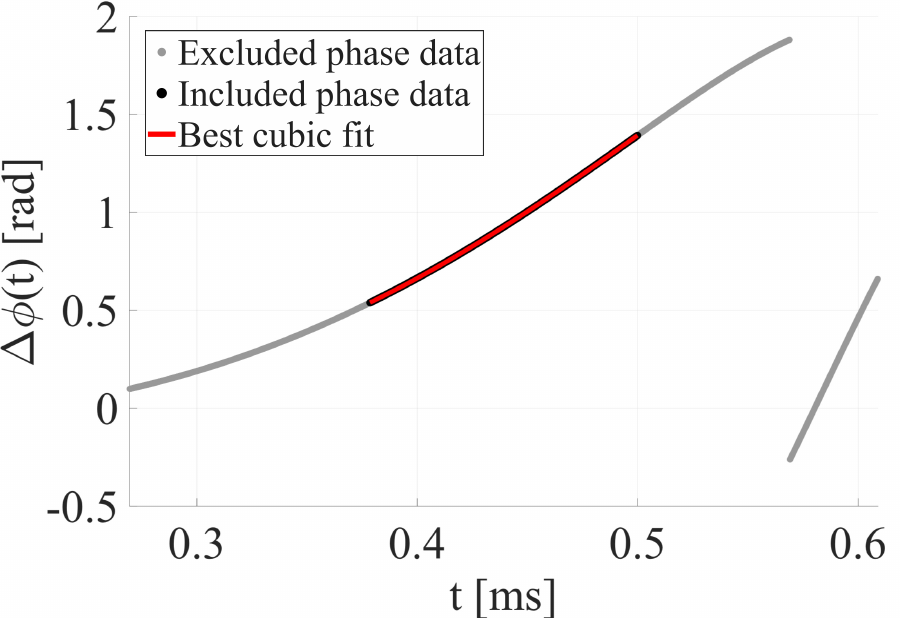}
  \caption{Relative global phase of the Airy main lobe interference fringe from the demodulated signal of the HPE – continuing the Ai-Ai collision scenario. The best cubic fit window choice, corresponding to the pink dot of Fig.~\ref{fig:ai-ai-heatmap}, is drawn in red relying on the highlighted data in black. The discontinuity near $t\approx0.6\,\mathrm{ms}$ is an extraction artefact marking the end of the contiguous main-lobe fringe. Past it, the linear- and quadratic-in-position (sub-cubic) parts of $\Delta\Phi$ vary from one fringe to the next, so the local plane-wave demodulation, whose carrier $k_\mathrm{f}(\tau)$ and centre $\xi_\mathrm{c}(\tau)$ are calibrated to a single fringe, cannot stitch the fragments into a continuous curve, and the unwrapped phase jumps. Only the contiguous segment before the jump enters the fit, so $c_3$ is unaffected.}
  \label{fig:ai-a-phase-best-fit}
\end{figure}

The best-fitting choice in Fig.~\ref{fig:ai-a-phase-best-fit} exhibits a slight cubic curvature.
Compared to Fig.~\ref{fig:airy_collision}, the preferred fit window does not cover the full temporal range of the interference fringe of the two main lobes. 
This can be attributed to the growing influence of the second-largest interference fringe, extending from $0.53\,\mathrm{ms}$ to $0.63\,\mathrm{ms}$ as an area of appreciable magnitude with a certain spatial width, hence affecting the measured ROI.   

Within the considered time window $t \in [0,1] \, \mathrm{ms}$ regarding the best fit coefficient in Table~\ref{tab:cubic-fit-coeffs}, the contributions $c_it^i$ of all polynomial orders are of comparable magnitude, except for the rather small cubic contribution, so the phase dynamics cannot be described as purely cubic. Nevertheless, the fit requires a non-zero cubic term $c_3$ with high statistical significance.
In general, the coefficients are well-determined considering the associated uncertainties which are reasonably small.

By construction, the estimator is debiased at $g=0$, so the extracted cubic coefficient matches the theoretical value within one-sigma uncertainties for fit intervals chosen along the valley. This agreement is therefore not itself a result but a consistency check of the calibration, and the relevant information lies not in the mean value of the top-ranked window measurements along the valley but in the structure of the robustness heatmap.
Only the cubic coefficient is compared to its theoretical prediction: the lower-order coefficients are strongly influenced by the demodulation, which subtracts the local linear term, and are therefore treated as nuisance parameters.
In particular, the model predicts a vanishing quadratic coefficient, whereas the fitted $c_2$ is significantly nonzero, an artifact of the approximate extraction rather than a failure of the model.
\renewcommand{\arraystretch}{1.2}
\begin{table}[h]
  \centering
  \caption{Best-fit coefficients for the Ai-Ai phase dynamics from the HPE,
  corresponding to the pink dot of Fig.~\ref{fig:ai-ai-heatmap}.
  The last column shows the order of magnitude of the phase contributions $|c_i t^i|$ at $t=0.10\,\mathrm{ms}$.
Numbers in parentheses denote $1\sigma$ uncertainties in the last digits
(e.g. $-3.47(3)\times10^{10}\equiv (-3.47\pm0.03)\times10^{10}$).}
  \begin{tabular}{llll}
    \hline\hline
    Coefficient & Value $\pm$ Standard Error  & Unit 
                & $|c_i t^i|$ [rad] \\
    \hline
    $c_3$ & $-3.47(3)\times10^{10}$   & $\mathrm{rad/s^3}$ & $10^{-2}$ \\
    $c_2$ & $5.67(4)\times10^{7}$     & $\mathrm{rad/s^2}$ & $10^{-1}$ \\
    $c_1$ & $-2.252(18)\times10^{4}$  & $\mathrm{rad/s}$   & $10^{0}$  \\
    $c_0$ & $2.826(26)$               & $\mathrm{rad}$     & $10^{0}$  \\
    \hline\hline
  \end{tabular}
  \label{tab:cubic-fit-coeffs}
\end{table}

When performing a third-order polynomial fit, the cubic coefficient is particularly sensitive to the selected input data, since the fitted range here lies within a region where the cubic term exhibits an inverted gradient compared to the dominant global trend.
This sensitivity becomes especially pronounced when only a few data points are available due to the finite temporal extent of the interference fringe.
In this case, a symmetric selection of data points around the inflection point is crucial, avoiding an overweighting of one curved tail of the fitted region.
Any asymmetry disturbs the fitting process relying on the essential underlying determination of the curvature, which is therefore directly reflected in $c_3$.
Thus, the prevention of this effective bias in the extracted cubic coefficient is important.

The hypersensitivity of the fitted cubic $c_3$ is illustrated in Fig.~\ref{fig:ai-ai-heatmap}, where high accuracy with respect to the theoretical value $c_{\mathrm{3,theo}}$ is achieved only for specific symmetrical pairs of start and end times.
The necessity of such symmetrized data selection is evident from the negative slope of the minima in the deviation from the theoretical $c_{\mathrm{3,theo}}$.
Accordingly, for a fixed $t_{\mathrm{start}}$, a larger $t_{\mathrm{end}}$ adds more data points to the right tail, 
hence increasing the leverage of late-time points, biasing the estimate toward smaller $c_3$ for a globally more negative trend of the phase, and vice versa.

\begin{figure}[h]
  \centering
  \includegraphics[
    width=\columnwidth,
    trim=25 125 95 375,clip  
  ]{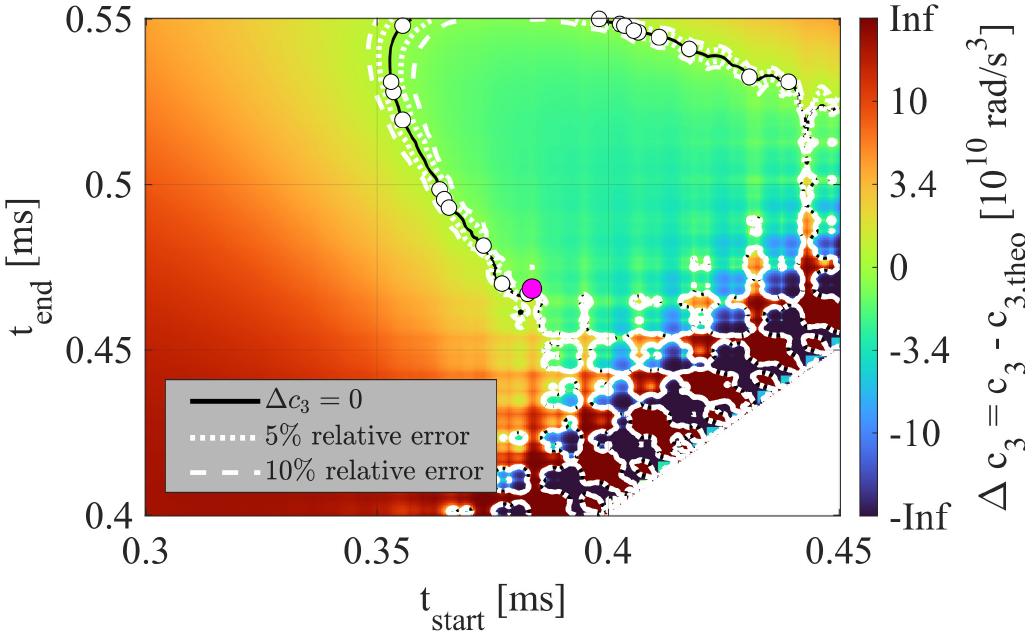}
  \caption{
    Deviation of $\Delta c_3 = c_3 - c_{\mathrm{3,theo}}$ across $[t_{\mathrm{start}}, t_{\mathrm{end}}]$
    for the Ai–Ai collision scenario, based on the DPE.
    As in Fig.~\ref{fig:ai-ai-heatmap}, the axes parameterize the fit window, 
    the white contour lines indicate relative errors of 5\% (dotted) and 10\% (dashed), 
    and the highlighted points mark the best discrete window combinations, with the pink dot denoting the closest value to $c_{\mathrm{3,theo}}$. 
    The region of small fit interval choices oscillates and diverges due to fit instability. 
    The colour scale uses the same signed, nonlinearly compressed mapping of $\Delta c_3$ as in Fig.~\ref{fig:ai-ai-heatmap}.}
  \label{fig:ai-ai-heatmap-density}
\end{figure}

For comparison with HPE, the experimentally simpler density-based phase extraction (DPE) is applied in the following, where $\sigma_{\mathrm{ROI}}$ is set to $0.34$ for an optimal extraction. 
The heat map in Fig.~\ref{fig:ai-ai-heatmap-density} lets one recognize a similar valley structure despite an enlarged region of fit instability of the fit in the lower right section. This divergence region for small fit interval choices appears to be method-inherent, since it does barely vanish for optimized parameter choices shown in Fig.~\ref{fig:ai-ai-heatmap-density}. 
Surprisingly, the upper area does not diverge to high negative deviations from $c_{\mathrm{3,theo}}$, but is restricted by the valley structure returning in form a u-shaped curve. 
This behavior appears to be linked to the choice of $\sigma_{\mathrm{ROI}}$ and could be explored further by a parameter scan.
Moreover, for larger fitting windows the contour lines are less tightly packed around the minima, as they are overall for the HPE,
possibly explained by the fact that more data tend to cause a more stable and robust fit with respect to the fit window selection.
Overall, the HPE distinguishes itself from the DPE by its higher accuracy also evident from the differences of the two heat scales.        
The best-fit choice has a very similar location with $t_{\mathrm{start}}\approx 0.38 \ \mathrm{ms}$, $t_{\mathrm{end}}\approx 0.45 \ \mathrm{ms}$.

When using the two phase-extraction methods, the best parameter choices differ: 
The HPE achieves optimal robust and precise $c_3$ estimates 
for very small $\sigma_{\mathrm{ROI}}$, which is reasonable since it directly accesses the
reconstructed amplitudes and works best with only few spatial data points that                 
carry the phase information. This corresponds to a narrow envelope, avoiding
large $\sigma_{\mathrm{ROI}}$ that would include 'noisier' regions further away
from the peak of the interference fringe, where the local plane-wave approximation becomes increasingly inaccurate. 
In contrast, the DPE works at its optimum for larger $\sigma_{\mathrm{ROI}}$,
because the spatial averaging of $\rho_{\mathrm{mean}}$ in $S_{\mathrm{DPE}}(\tau)$ requires a sufficiently large area to efficiently
cancel out local fluctuations and small-scale variations, and to average over noise, in order to provide the background which is subtracted from the pure density to obtain the evolution in Eq. 26.
However, this larger Gaussian window introduces the influence of neighbouring interference fringes as a
systematic bias
and also decreasing the effective quality in terms of the signal-to-noise ratio of the extracted phase signal, which affects the robustness-map structure and increases the error bars of the fit coefficients, as seen in
Table~\ref{tab:cubic-fit-coeffs-density}. Therefore, the parameters are
adjusted separately for each method to minimize variance and maximize
stability.

Furthermore, the combined effect of the bias from adjacent fringes entering as a systematic error, and the local plane-wave approximation used to define the carrier leads to DPE uncertainties that are systematically larger than, or at best comparable to, the HPE benchmark. In particular, the minimal DPE uncertainty observed in the robustness map does not fall below the corresponding HPE value.

\begin{figure}[h]
  \centering
  \includegraphics[width=\columnwidth,
  trim=25 10.8 98 440,clip]{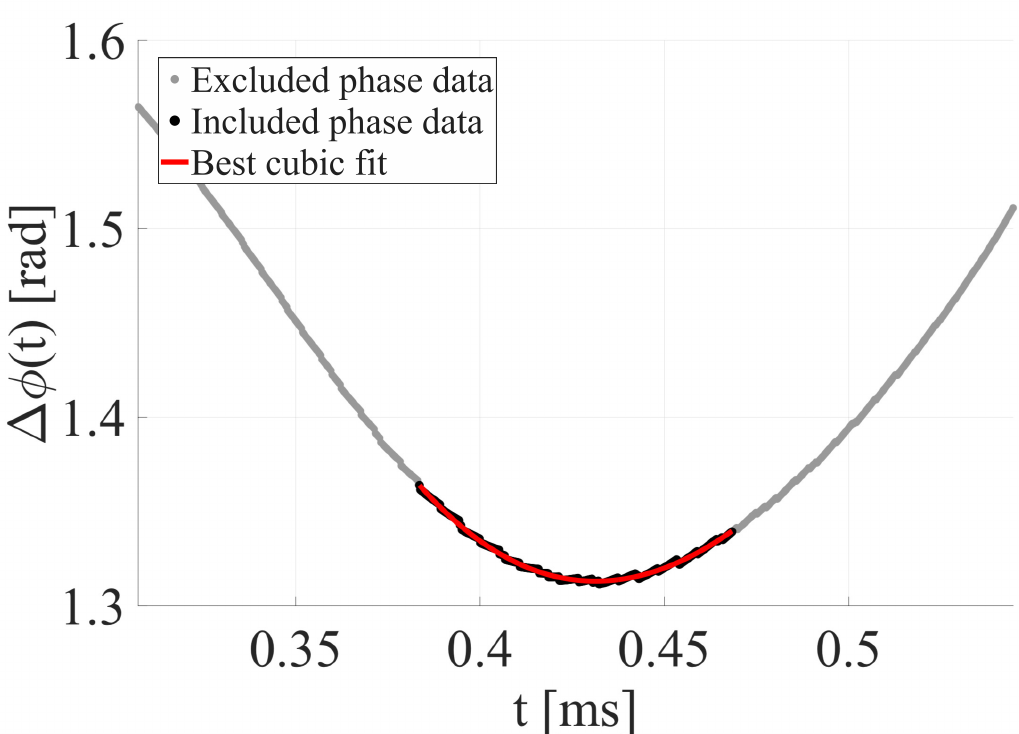}
  \caption{
  Relative global phase of the Airy main lobe interference fringe from the demodulated DPE signal for the Ai–Ai collision scenario.
  The best cubic fit window is shown in red and is based on the highlighted data points in black. 
  This interval falls within the cubic regime, where the local slope reverses relative to the global trend.}
  \label{fig:ai-a-phase-best-fit-density}
\end{figure}

In addition, another major difference is that the best-fit for DPE in Fig.~\ref{fig:ai-a-phase-best-fit-density} is not centered around the inflection point, assuming the underlying phase dynamics from HPE in Fig.~\ref{fig:ai-a-phase-best-fit} provide a clear, extracted signal as a good approximation of the true phase dynamics governing also this second case.
Particularly since the fitted data are not very similar, which may be due to the fundamentally different phase extraction methods, a deeper examination would be necessary to infer conclusions with high confidence and to differentiate between various contributions.
Fig.~\ref{fig:ai-a-phase-best-fit-density}, in combination with the evolution in Fig.~\ref{fig:airy_collision}, suggests that spatially well-separated data with a clear distance to other interference patterns are favored in order to obtain the most pure signal and therefore constraining possible fit windows, since the best fit time interval does not cover the whole interference fringe - otherwise, the curvature of the phase dynamics is distorted.
  
The DPE-fit coefficients show significantly larger uncertainties, while the cubic term still agrees with the theoretical value.
In particular, the standard errors are increased by about one order of magnitude,
which originates in the fundamentally less precise extraction technique involving additional averaging and approximations.
In general, this density-based method is still an elegant approach to access information about the phase dynamics with reasonable output measurements.
\begin{table}[h]
  \centering
  \caption{Fit coefficients of Ai-Ai phase dynamics (DPE) at the best-fit window, corresponding to the pink dot in Fig.~\ref{fig:ai-ai-heatmap-density}.
  The last column shows the order of magnitude of the phase contributions $|c_i t^i|$ at $t=0.10\,\mathrm{ms}$.}
  \begin{tabular}{llll}
    \hline\hline
    Coefficient & Value $\pm$ Standard Error  & Unit & $|c_i t^i|$ [rad] \\
    \hline
    $c_3$ & $-3.5(4)\times10^{10}$   & $\mathrm{rad/s^3}$ & $10^{-2}$ \\
    $c_2$ & $6.6(5)\times10^{7}$     & $\mathrm{rad/s^2}$ & $10^{-1}$ \\
    $c_1$ & $-3.72(21)\times10^{4}$  & $\mathrm{rad/s}$   & $10^{0}$  \\
    $c_0$ & $7.94(29)$               & $\mathrm{rad}$     & $10^{0}$  \\
    \hline\hline
  \end{tabular}
  \label{tab:cubic-fit-coeffs-density}
\end{table}

\subsection{Cubic Airy--Gaussian Phase}

\begin{figure}[h]  
  \centering
  \includegraphics[width=\columnwidth, trim=5 20 190 0,clip  
  ]{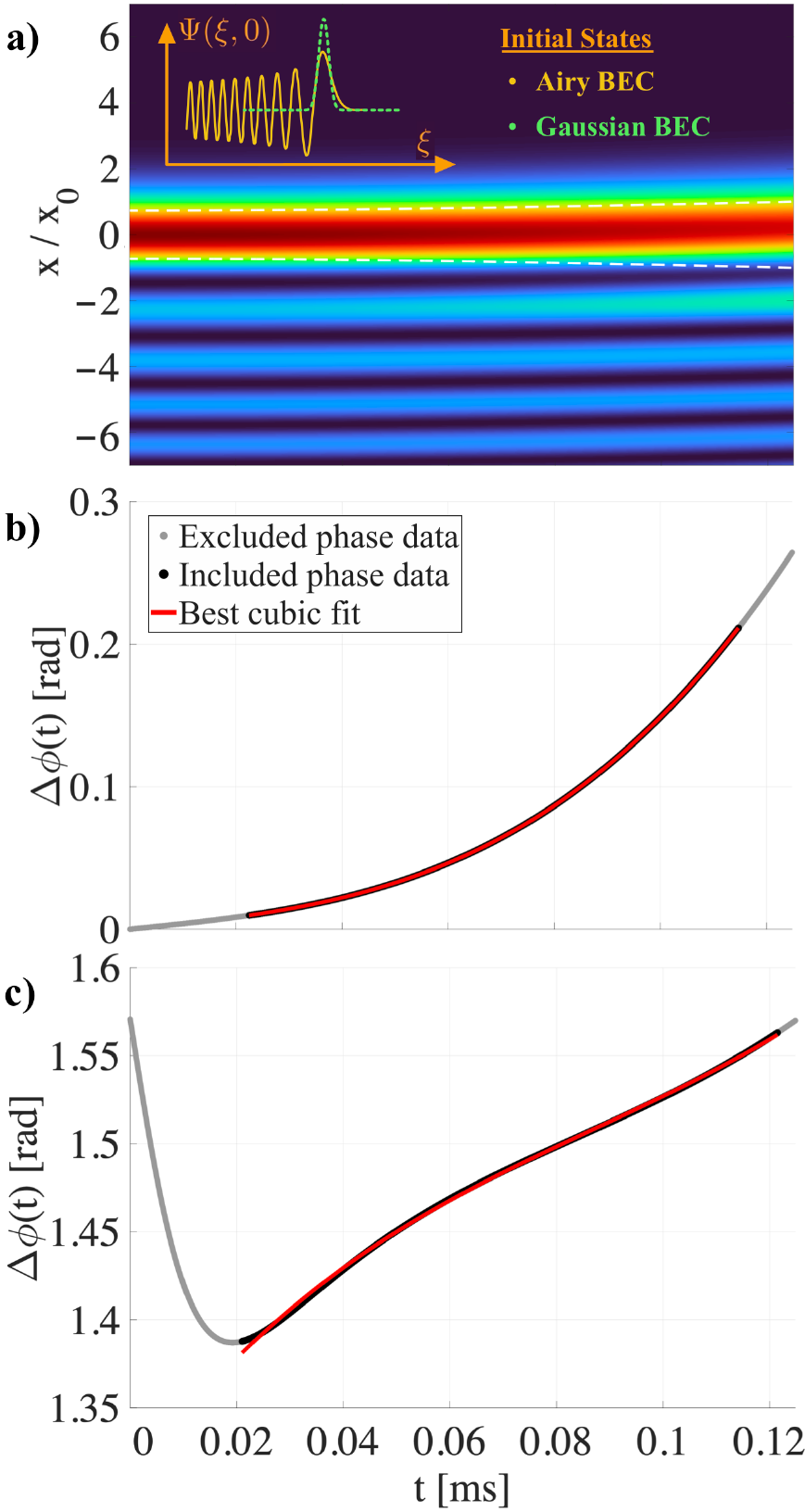} 
  \caption{Time evolution of probability density profile $|\Psi(x,t)|^2$ in free space, shown in a): Interference of an Ai-BEC and a Gaussian wave packet creating an initial overlap with the main lobe of the Airy profile. The white contour line marks the evolving one-sigma-width of the dispersing Gaussian packet. \newline
  Relative global phase of the overlap region from the demodulated b) HPE and c) DPE signal for the Ai–G collision.
  The best cubic fit window is shown in red and is based on the black data points.}      
  \label{fig:ai-g-collision-merge}
\end{figure}
\begin{figure}[h]
  \centering
  \includegraphics[
    width=\columnwidth,
    trim=35 215 80 15,clip  
  ]{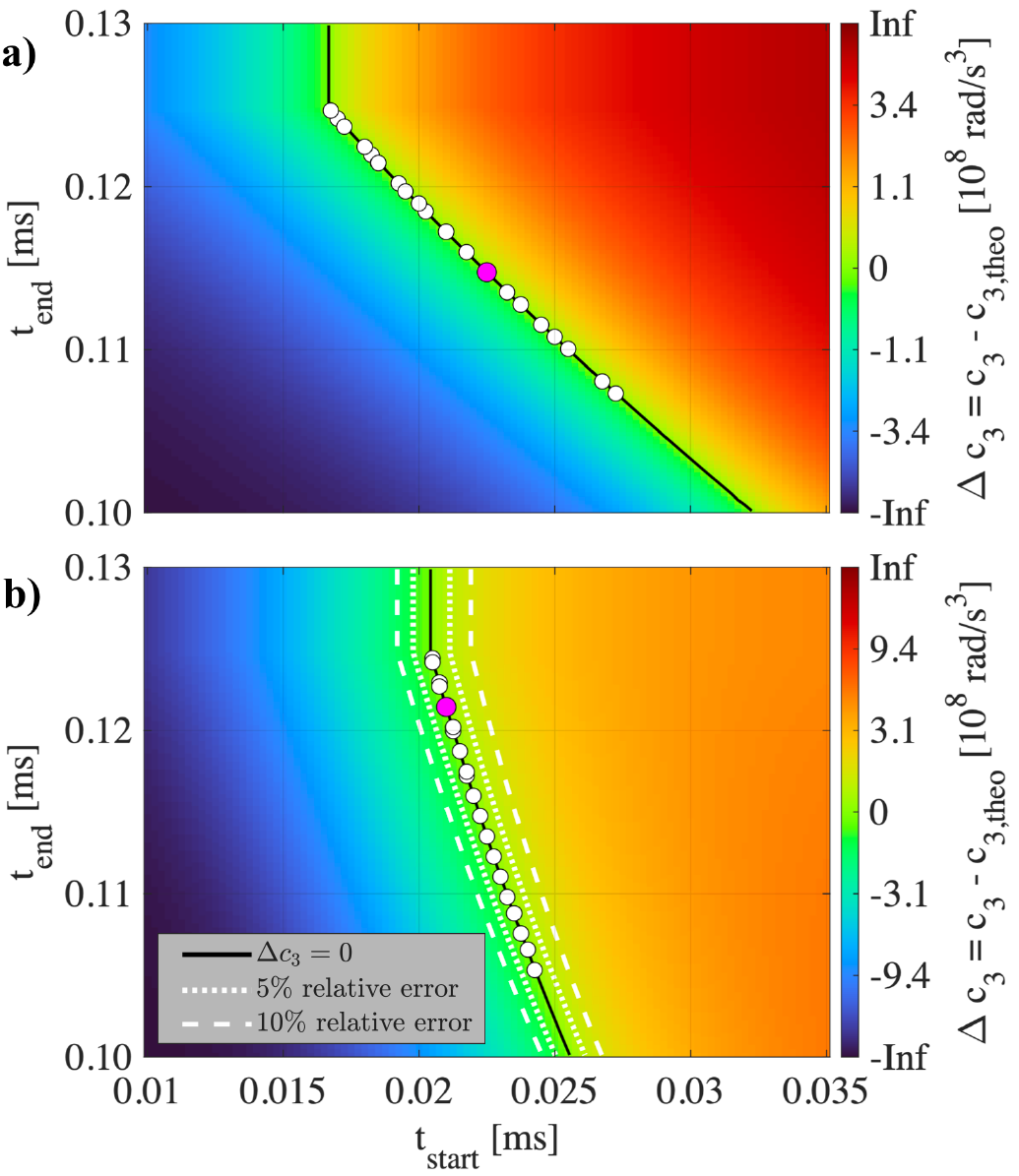}
  \caption{Deviation of $\Delta c_3 = c_3 - c_{\mathrm{3,theo}}$ across $[t_{\mathrm{start}}, t_{\mathrm{end}}]$
    for the Ai–G collision, based on the a) HPE and b) DPE.
    As in Fig.~\ref{fig:ai-ai-heatmap}, the axes parameterize the fit window, 
    the white contour lines indicate relative errors of 5\% (dotted) and 10\% (dashed), 
    and the highlighted points mark the best discrete window combinations, with the pink dot denoting the closest value to $c_{3,\text{theo}}$.
    The colour scale uses the same signed, nonlinearly compressed mapping of $\Delta c_3$ as in Fig.~\ref{fig:ai-ai-heatmap}.}
  \label{fig:ai-g-heatmap-merge}
\end{figure}

 
In Fig.~\ref{fig:ai-g-collision-merge} a sharply peaked Gaussian BEC is superimposed with the main lobe of an Ai-BEC.
For analyzing the collision, $\sigma_{\mathrm{ROI}}$ is set to $0.20$ for HPE and $0.21$ for DPE, 
because the spatial extension of the Gaussian wave packet has to be covered well-enough in the measurement.
Both robustness maps in Fig.~\ref{fig:ai-g-heatmap-merge} exhibit a noiseless valley structure similar to the HPE-based Ai-Ai scenario, which is the closest simulated description to the theory of the Ai-Ai case here available.
Moreover, there is a sharp bend upwards in the valley structure in the vertical direction for both methods.
For HPE fitting windows with $t_{\mathrm{start}}\approx 0.017 \ \mathrm{ms}$, there appear to be several best choices for $t_{\mathrm{end}}$ 
starting with the lowest at $0.125 \ \mathrm{ms}$ till the boundary of the heatmap.
A similar behavior with slightly different values is observed for the DPE case.
In combination with the slightly decreased slope of the minima valleys for HPE and an apparently reduced slope for DPE, this might suggest avoiding early measurement data but allowing a rather generous choice of the final boundary.
Especially, the robustness is remarkable since for both extraction methods large regions of the heatmaps show deviations in the order of $10^8$ - equating approximately three orders smaller than the theoretical value $c_{\mathrm{3,theo}}=1.161642 \times 10^{11} \ \mathrm{rad/s^3}$. 
The best fit simulation measurement is $(1.16164 \pm 0.00009)\times 10^{11} \ \mathrm{rad/s^3}$ for HPE and $(1.162 \pm 0.019)\times 10^{11} \ \mathrm{rad/s^3}$ for DPE with the smallest deviations to the theoretical value.
The other top 19 values are seemingly equally distributed for HPE and DPE except for short fitting intervals, requiring again a minimum level of data for an accurate fit, and early time inclusion.
For HPE, the contour lines are not even drawn, and for DPE they enclose a noticeable area around the minima valley.

Regarding the relative phase data and their best fit windows in Fig.~\ref{fig:ai-g-collision-merge} b) and c), 
the two extraction methods share the same general ascending, nonlinear trend. 
However, the DPE phase data start with an irregular peak for early times.
Both best fit choices use a time interval starting approximately at $t_{\mathrm{start}} = 0.02 \ \mathrm{ms}$, and hence DPE excludes this artefact.       

The cubic coefficients of Table~\ref{tab:cubic-fit-coeffs-heterodyne-ai-g} and \ref{tab:cubic-fit-coeffs-density-ai-g} agree with $c_{\mathrm{3,theo}}= 1.161642 \times 10^{11} \ \mathrm{rad/s^3}$ within the one-sigma uncertainties.
The HPE suggests a strong dominance of the cubic term over its nuisance parameters, due to the enhanced cubic contributions in comparison to the Ai-Ai case.
The reason lies in the asymmetry of the collision: in the Ai-G phase difference (Eq.~21) the Airy and Gaussian cubic-in-time terms carry the same sign and add, whereas in the Ai-Ai case (Eq.~19) the two intrinsic Airy cubic terms enter with opposite sign and nearly cancel for near-symmetric packets.
Therefore, an excessive, dominant role of the cubic phase can be achieved – here by approximately one order of magnitude over the linear term and two orders over the quadratic term for the demodulated, fitted phase data, enabling the HPE observer to clearly differentiate between the polynomial contributions and accessing the cubic phase coefficient.
For DPE in Table~\ref{tab:cubic-fit-coeffs-density-ai-g}, the contributions are not as divergent as for HPE, potentially because of the concave artefact bulge in the phase data at $t=0.05 \mathrm{ms}$, but still in the same order – neglecting the irrelevant offset, such that the cubic contribution is well expressed to be extracted.

Summarizing the emergent picture, it can be concluded that rather small differences for the Ai-G collision across the different techniques, HPE and DPE, can be noticed, in contrast to the sharp distinction to the Ai-Ai scenario based on the asymmetry in the phase dynamics of Ai-G.

\begin{table}[h]                    
  \centering
  \caption{Fit coefficients of Ai-G phase dynamics (HPE) at the best-fit window, corresponding to the pink dot in Fig.~\ref{fig:ai-g-heatmap-merge}~a). The last column shows the order of magnitude of the phase contributions $|c_i t^i|$ at $t=0.10\,\mathrm{ms}$.}
  \begin{tabular}{llll}
    \hline\hline
    Coefficient & Value $\pm$ Standard Error  & Unit & $|c_i t^i|$ [rad] \\
    \hline
    $c_3$ & $1.16164(9)\times10^{11}$  & $\mathrm{rad/s^3}$ & $10^{-1}$ \\
    $c_2$ & $-8.173(18)\times10^{5}$   & $\mathrm{rad/s^2}$ & $10^{-3}$ \\
    $c_1$ & $4.1152(12)\times10^{2}$   & $\mathrm{rad/s}$   & $10^{-2}$ \\
    $c_0$ & $-3.938(22)\times10^{-4}$  & $\mathrm{rad}$     & $10^{-4}$ \\
    \hline\hline
  \end{tabular}
  \label{tab:cubic-fit-coeffs-heterodyne-ai-g}
\end{table}

\begin{table}[h]
  \centering
  \caption{Fit coefficients of Ai-G phase dynamics (DPE) at the best-fit window, corresponding to the pink dot in Fig.~\ref{fig:ai-g-heatmap-merge}~b). The last column shows the order of magnitude of the phase contributions $|c_i t^i|$ at $t=0.10\,\mathrm{ms}$.}
  \begin{tabular}{llll}
    \hline\hline
    Coefficient & Value $\pm$ Standard Error  & Unit & $|c_i t^i|$ [rad] \\
    \hline
    $c_3$ & $1.162(19)\times10^{11}$ & $\mathrm{rad/s^3}$  & $10^{-1}$ \\
    $c_2$ & $-3.26(4)\times10^{7}$   & $\mathrm{rad/s^2}$  & $10^{-1}$ \\
    $c_1$ & $4.382(22)\times10^{3}$  & $\mathrm{rad/s}$    & $10^{-1}$ \\
    $c_0$ & $1.3034(4)$              & $\mathrm{rad}$      & $10^{0}$  \\
    \hline\hline
  \end{tabular}
  \label{tab:cubic-fit-coeffs-density-ai-g}
\end{table}

In contrast to the Ai-Ai collision, the generally shown time evolution of interest is much smaller due to the rapid dispersion of the peaked Gaussian packet, 
but the length of the fitted time interval is in a similar order of $t_{\mathrm{fit-interval}} \approx 0.1 \ \mathrm{ms}$ for both types of collisions.

In addition, the Ai-G phase extraction distinguishes from the Ai-Ai algorithm by eventually subtracting a geometric correction $\phi_{\mathrm{geom}}(\tau) = k_\mathrm{f}(\tau) \, \xi_\mathrm{c}(\tau)$ of the refined phase to obtain the corrected phase
\begin{equation}
    \phi_{\mathrm{corr}}(\tau) = \phi_{\mathrm{rel}}(\tau) - \phi_{\mathrm{geom}}(\tau) \ .              
\end{equation}
This term originates in the centered demodulation, whose purpose is to make the extraction robust to small errors in the determination of $k_\mathrm{f}$, in Eq. 22 and 28.
An uncentered demodulation could cause phase ramps $e^{-\mathrm{i}(k_{\mathrm{f,true}} \ + \ \Delta k_\mathrm{f} )\xi}$ across the Gaussian window, which would enter as a systematic bias in the measured amplitude and phase.
The spatially centered method provides, in case of a slightly misdetermined $k_\mathrm{f}$, a symmetrical ramp which mostly cancels out.
For the Ai-Ai case, the evolution, apart from the small difference of the Airy length scales, is highly symmetric, and the interference fringe is quasi-stationary in Fig.~\ref{fig:airy_collision}, resulting in an almost constant $\xi_\mathrm{c}(\tau) = \xi_{\mathrm{c,fringe}}$.
Since the main-lobe wavenumber is set by the linear-in-$\xi$ part of the accumulated phase, $\partial_\xi\Phi_{Ai}=\tau/2$,\footnote{The main-lobe wavenumber follows from the accumulated phase of the evolved packet, $\arg\Psi_{Ai}=\Phi_{Ai}+\arg[\mathrm{Ai}(\xi-\tau^2/4+ia\tau)]$. In the small-truncation limit the Airy envelope is real and contributes no phase gradient between its zeros, so at the main lobe $\partial_\xi\arg\Psi_{Ai}=\partial_\xi\Phi_{Ai}=\tau/2$, linear in $\tau$, up to an $\mathcal{O}(a\tau)$ correction that vanishes as $a\to0$, where the Airy argument becomes real. The identification of this gradient with the numerically extracted carrier is justified in the extraction subsection above.} the differential carrier $k_\mathrm{f}(\tau)$ grows only linearly in time, and with $\xi_\mathrm{c}\approx\mathrm{const}$ the geometric factor $k_\mathrm{f}(\tau)\,\xi_\mathrm{c}(\tau)$ is likewise linear in time. It therefore enters only the nuisance linear coefficient, leaving the cubic coefficient $c_3$ unaffected, so that no major correction is required.
In the asymmetric Ai-G scenario, by contrast, $\xi_\mathrm{c}(\tau)$ moves as the self-accelerating Airy main lobe sweeps across the dispersing Gaussian, so that $k_\mathrm{f}(\tau)\,\xi_\mathrm{c}(\tau)$ acquires a higher-order time dependence that would bias $c_3$ and therefore has to be corrected.
Furthermore, for consistency with the cubic phase model, the higher-than-third-order remainder of the Gaussian Gouy term
$-\frac{1}{2}\!\left[\arctan(\eta)-\left(\eta-\frac{\eta^3}{3}\right)\right]$ is subtracted from $\phi_{\mathrm{corr}}(\tau)$ prior to fitting.
Unlike the Ai-Ai case, the Ai-G phase does not exhibit a local slope opposite to the overall trend within the fit window, which also makes the fitting procedure significantly more robust, i.e. far less sensitive to window-dependent bias in the estimator, as seen in the heatmaps.                  

For Ai-Ai, a slight asymmetry in the Airy length scale scales between the two Airy packets is introduced to avoid an exact cancellation of the relative cubic phase.
Nevertheless, the cubic phase contribution in the highly symmetric Ai-Ai collision remains small compared with lower-order terms, such that finite-window limitations and extraction systematics make isolating the cubic dynamics comparatively fragile.
This highlights a trade-off in the collision design: a more symmetric Ai-Ai configuration improves cancellation of nuisance contributions and makes geometric phase corrections largely irrelevant, but it also suppresses the net cubic signal.
In contrast, the intentionally asymmetric Ai-G collision yields a dominant cubic phase contribution, which substantially improves robustness and parameter separability in the polynomial fit.

The major difference between the two constructed collisions lies in their robustness, i.e. in the reduced window sensitivity of the approximately debiased estimator, and in the greater precision of the cubic coefficient, primarily owing to its dominant contribution for Ai-G. 
The overall significantly higher robustness of the Ai-G collision - in particular two orders of magnitude less sensitive to the fit interval selection, and the congruence of both Ai-G robustness maps of the HPE and the DPE, in comparison with the obvious differences for Ai-Ai, recommend the Ai-G case with DPE as the preferred simplest scenario to study, again reminding of the assumption that the HPE method is the technique providing the closest measurement describing the corresponding analytically derived theoretical model.                                                                    
A further advantage is the higher precision for Ai-G. 
Despite a more than twice larger $c_{\mathrm{3,theo}}$ of Ai-G compared to Ai-Ai, the relative as well as the absolute standard deviations are significantly smaller for Ai-G-HPE and Ai-G-DPE,
which can be explained by the not only more window-robust, but enhanced stable fit due to the dominant cubic phase, corresponding to the asymmetry of the collision effecting the phase dynamics. 
For instance, the experimentally preferred DPE yields standard errors of $\Delta c_{\mathrm{3, Ai-G}} = 1.9 \times 10^9 \ \mathrm{rad/s^3}$ and $\Delta c_{\mathrm{3, Ai-Ai}} = 4 \times 10^9 \ \mathrm{rad/s^3} \ \approx \, 2 \cdot  \Delta c_{\mathrm{3, Ai-G}}$, equating to the relative errors of $1.64 \,  \%$ for Ai-G and $11.43 \, \%$ for Ai-Ai.

\subsection{Uncertainty Estimation}
Rather than relying on a single optimal fit window, for each dataset one can average over a small family of nearby window points centered on the robustness valley and apply the heteroskedasticity- and autocorrelation-consistent (HAC) Newey-West estimator that accounts for the potentially strong correlations induced by overlapping windows in the averaging procedure. 
The reported errors should be interpreted as conservative, data-driven, systematics-aware uncertainty estimate. Technical details of the construction of the estimator are given in Appendix~\ref{app:uncertainty}.
Additional validation of the extracted $c_3$ scaling is provided in Appendix~\ref{app:airysweep}.

\subsection{Nonlinearity Influence on Cubic Dynamics}

\begin{figure}[h]
  \centering
  \includegraphics[width=\columnwidth,
    trim=15 205 60 200.5,clip  
  ]{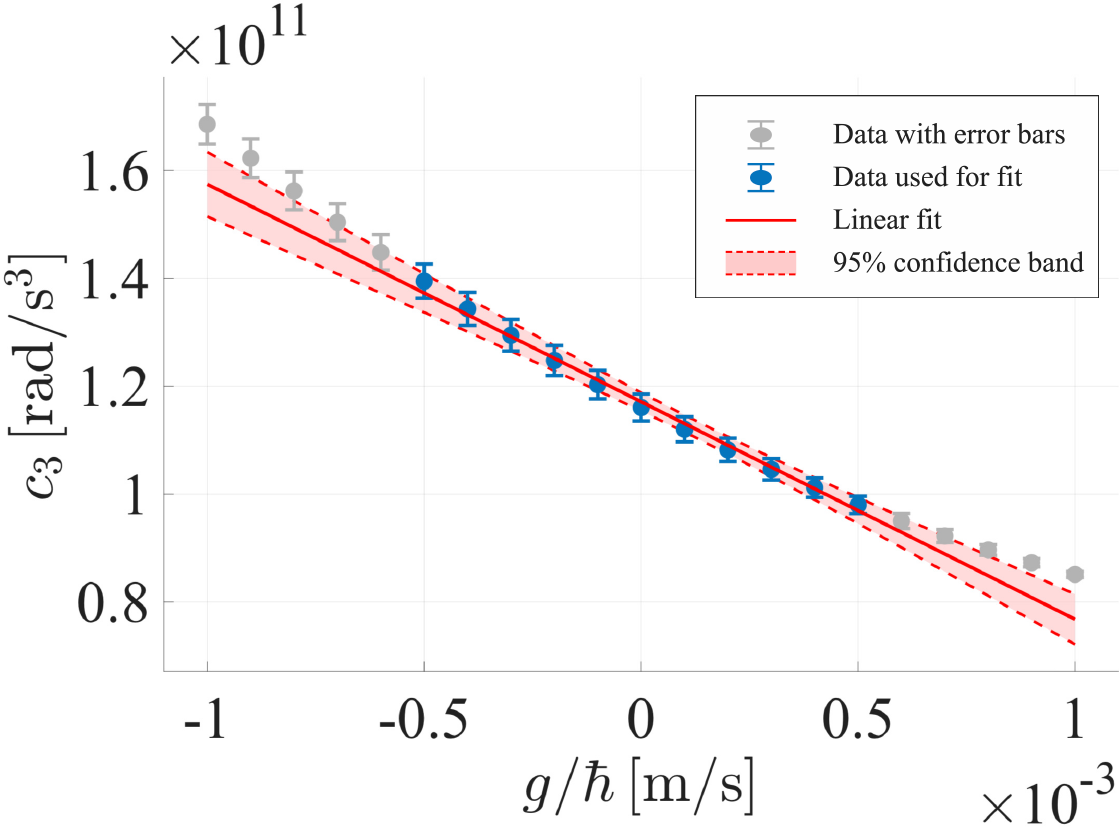}
  \caption{
    Cubic coefficient $c_3(g)$ as a function of the nonlinearity parameter $g/\hbar$, extracted from the Ai-G scenario using DPE with HAC Newey-West uncertainty and mean estimation, and more precisely $\hat c_3(g)$ as an approximately debiased estimator calibrated via the analytical linear baseline for $g=0$. The blue-highlighted subset indicates the perturbative range used for the linear approximation. The red confidence band is bounded by the dashed red lines.}
  \label{fig:c3vsg}
\end{figure}

Eventually, the influence of the scaling factor $g$ of the nonlinearity on the cubic coefficient $c_3$ can be studied.
In particular, $c_3(g)$ is extracted from the Ai--G collision using the DPE. The estimator $\hat c_3(g)$ is approximately debiased via calibration by the analytical linear baseline g=0, and uncertainties are estimated via the HAC Newey-West procedure to obtain a conservative, systematics-informed error level.
The $g=0$ calibration removes the $g$-independent part of the extraction bias, but not biases that the nonlinear dynamics, such as self-focusing and defocusing, may introduce at $g\neq0$. These are assumed negligible across the perturbative window, consistent with the residual analysis below, rather than eliminated, so the reported linear response is a controlled first-order estimate under this assumption.

Special attention is paid to experimentally motivated parameters: the rubidium-87 mass $m = 1.44 \times 10^{-25}\,\mathrm{kg}$ and the corresponding force of the eigenpotential for the initial Airy profile, $F = m\,a_{\mathrm{grav}}$ with $a_{\mathrm{grav}}= 9.81\,\mathrm{m/s^2}$ used in the simulation.

The central data point in Fig.~\ref{fig:c3vsg} corresponds to the calibration at $g=0$.
For $g<0$, the attractive, focusing interaction produces bright-soliton-like behavior, whereas for $g>0$ the repulsive, defocusing one leads to dark-soliton-like density notches.

In the weakly nonlinear regime, the cubic coefficient $c_3(g)$ is a smooth function of the interaction, which enters the Gross--Pitaevskii evolution analytically, so that its small-$g$ behavior is governed by the leading term of a Taylor expansion,
$c_3(g)=c_3(0)+ \alpha \,(g/\hbar)+\mathcal{O}\big(g^2\big)$,
the standard first-order perturbative response. The analysis is therefore restricted to a narrow window around $g=0$.
The perturbative window is chosen as the largest symmetric interval around $g=0$ in which the linear model remains consistent with the data, the residuals staying within their uncertainties and free of systematic curvature. Here the cubic coefficient is well described linearly in $g$ for $g/\hbar \in [-0.5\times 10^{-3},\,0.5 \times 10^{-3}]\,\mathrm{m/s}$.\footnote{For an order-of-magnitude context, the per-atom one-dimensional coupling is $g_{1\mathrm{D}}/\hbar \simeq 2\,\omega_\perp a_\mathrm{s}$ (in $\mathrm{m/s}$, for $a_\mathrm{s}\ll a_\perp$), so that the effective control parameter is $g_\mathrm{eff}/\hbar = N\,g_{1\mathrm{D}}/\hbar = 2N\,\omega_\perp a_\mathrm{s}$. For rubidium-87 ($a_\mathrm{s}\approx 5.3\,\mathrm{nm}$, $\omega_\perp/2\pi\approx 5$--$30\,\mathrm{kHz}$) the per-atom value is $g_{1\mathrm{D}}/\hbar\sim 10^{-4}$--$10^{-3}\,\mathrm{m/s}$~\cite{bouchoule,vanEs,Proukakis2006}, for a condensate of $N\sim10^{3}$--$10^{4}$ atoms the perturbative window $|g_\mathrm{eff}/\hbar|\le 0.5\times 10^{-3}\,\mathrm{m/s}$ is then reached by tuning $a_\mathrm{s}$ close to zero. As the coupling $g_\mathrm{eff}/\hbar = 2N\,\omega_\perp a_\mathrm{s}$ is mass-independent~\cite{olshanii1998}, the same $(N,\omega_\perp,a_\mathrm{s})$ realize this window for other species as well, the rubidium-87 mass fixing only the representative physical scale ($x_0$, $c_3$, $\tau_0$). Tuning $a_\mathrm{s}$ well below the native rubidium value requires a broad Feshbach zero-crossing, as available in $^{39}$K or $^{133}$Cs rather than in $^{87}$Rb~\cite{chin2010feshbach}, the native scattering length otherwise placing the sample beyond the linear-response window, in the nonlinear regime.}
Within the shown range, the estimated uncertainty of $c_3$ decreases for larger $g$.

A weighted least-squares fit of the linear model $c_3(g)= \alpha \, g/\hbar+ \beta$ in the window $|g/\hbar|\le 0.5\times 10^{-3} \, \mathrm{m/s}$ with the weights $w_i=1/\sigma_i^2$ yields $\alpha = (-4.033 \pm 0.012)\times 10^{13}  \ \mathrm{rad \; m^{-1} \; s^{-2}}$ and $\beta = (1.171 \pm 0.008)\times 10^{11} \ \mathrm{rad/s^3}$ with standard errors under the simplifying assumption of independent point-to-point errors. \footnote{The pointwise uncertainties $\sigma_i$ are obtained from a HAC Newey-West estimate applied to the correlated time series underlying each averaged data point $c_3(g_i)$. The uncertainties are analytic, model-based standard errors due to the generalized least square fit. They are inferred from the residual scatter. No per-point phase uncertainties enter because phase noise is not explicitly modeled. In addition, the standard errors might ignore possible correlations among adjacent data points, so the fit is used as a phenomenological first-order approximation in the perturbative regime. The reduced chi-square obtained from the weighted fit is $\chi^2_\mathrm{red}=0.195$. Values below unity can indicate that the quoted $\sigma_i$ are conservative and/or that residual correlations between estimated points are not fully captured by a diagonal weighting. Therefore, $\chi^2_\mathrm{red}$ is primarily used as a qualitative consistency check and the assessment relies on the residual structure and fit-window stability to evaluate the adequacy of the linear approximation.} 

Thus, according to Fig.~\ref{fig:c3vsg} the necessary required precision for confident experimental confirmation of the prediction of the linear dependence of $c_3$ on $g$ is therefore a one-sigma-uncertainty strictly smaller than $0.2 \times 10^{11} \approx | c_3(g_{\mathrm{max, linear}}) -  c_3(0)|$.
Within the chosen window, the residuals remain well below the quoted uncertainties, all points lying within $\lesssim 1\sigma$, so the linear approximation provides a controlled first-order description of the response.
The residuals also exhibit a weak, even-in-$g$ curvature consistent with subleading $\mathcal{O}(g^2)$ corrections, which also lifts the fitted intercept above $c_{\mathrm{3,theo}}$ as a linear-model effect rather than a per-point bias, and the fitted slope $\alpha$ is therefore interpreted as the leading perturbative response coefficient of $c_3$ to the effective coupling $g_\mathrm{eff}=N\,g_{1\mathrm{D}}$ in the small-$|g|$ regime. Physically, this response may be read as a local mean-field modification of the eigenforce~\cite{pellner2026avs,pellner2026thesis}.
Beyond $|g/\hbar|>0.5 \times 10^{-3} \, \, \mathrm{m/s}$ the data become clearly nonlinear, the linear model is underparametrized, and higher-order corrections are required.

For the Ai-Ai case at stronger nonlinearity, the fringe pattern becomes dominated by mean-field effects: in particular, the phase extraction relying on the initial calibration for general debiasing requires an approximately constant fringe geometry, whose spatially fine interference structure however becomes progressively disturbed resulting in less accurate phase extraction until ultimately breaking down. \footnote{These intricacies  could be dealt with by clever intermediate calibration, which would open the future realm to the strong interaction regime for Ai-Ai.}
In contrast, the initial overlap of the Airy main lobe and the Gaussian packet in the Ai-G configuration – shown here – remains substantially more robust. This motivates using the Ai-G setup as a practical nonlinearity benchmark for probing the regime even beyond the perturbative limit.

\section{Conclusion}    
The interference of Airy-shaped Bose-Einstein condensate wave packets with each other (Ai-Ai) or with a Gaussian profile (Ai-G) in free space gives access to their intrinsic cubic-in-time phase through the main interference fringe.
Two physically feasible methods, based on heterodyne or on pure density measurements, resolve these peculiar phase dynamics, characteristic of the self-accelerating Airy packet \cite{berrybalasz1979}.
Therefore, the influence of the Gross-Pitaevskii-nonlinearity can be studied by adding a small perturbation in the conventional linear Schrödinger equation as an approximation of the Gross-Pitaevskii equation, which is implemented in the simulation via a split step method.
Within $|g/\hbar|\le 0.5\times 10^{-3}\,\mathrm{m/s}$, the cubic coefficient of the phase dynamics exhibits a linear dependence in $g/\hbar$.

The asymmetric Ai-G configuration enhances the cubic contribution, is significantly more robust to a fit-window-induced bias than the more fragile, near-symmetric Ai-Ai case, and
enables access to the cubic phase with high precision.
Within each collision, the heterodyne phase extraction is at least one order of magnitude more precise than the density-based phase extraction, which still yields reasonably small uncertainties and is experimentally easier to access.
Overall, Ai-G is about one order of magnitude more precise than Ai-Ai for the latter extraction technique.
The dependence of the cubic coefficient on the interaction thereby turns the Ai-G collision into an interferometric probe of weak mean-field nonlinearities, physically accessible from density observables.

The reported uncertainties follow from a heteroskedasticity- and autocorrelation-consistent Newey-West estimator, which accounts for the correlations induced among the overlapping fit windows and yields a conservative, data-driven uncertainty estimate. Since this estimator is constructed for correlated, noisy data, it extends directly once detection noise is included, so the present noise-free estimates constitute a controlled lower bound on the experimental uncertainty.\footnote{No detection noise is imposed here, as realistic shot noise, finite imaging resolution, and detection efficiency are system-dependent, to be complemented by a system-specific noise budget in a dedicated experimental study.}

In the previous analysis, the dynamics were reduced to an effective one-dimensional problem. In three dimensions, narrow Gaussian envelopes in the neglected directions should enable comparable collision geometries.
It is worth recalling that the force-induced cubic-in-time phase is not limited to the Airy profile, but emerges for arbitrary wave packets evolving in a linear potential \cite{kennard1927}, whose full cubic phase dynamics are cast in a physically interpretable closed form in a companion study \cite{pellner2026avs,pellner2026thesis}.
This phase is platform-independent, not restricted to ultracold atoms: any condensate governed to good approximation by effective Gross-Pitaevskii dynamics with an engineered potential gradient – including e.g. exciton-polariton \cite{Rozenman2022Dispersion,Rozenman2018LongRange}
or photonic Bose-Einstein condensate realizations \cite{Bloch2022PhotonBEC} – should in principle exhibit the same cubic-in-time phase and admit analogous interference-based phase reconstruction \cite{deng2010,carusotto2013}.

\subsection{Experimental Outlook}

Phase reconstruction via interference offers a direct pathway toward observing the intrinsic cubic phase of self-accelerating condensates in cold-atom experiments. 
By releasing a BEC prepared in an Airy-like state and overlapping it with a reference Gaussian cloud, one can image the interference fringes and extract the relative phase of the condensates. 
Such measurements would directly test the predicted cubic phase dynamics and provide an experimentally accessible observable that is sensitive to weak mean-field interactions. 
Tuning the interaction strength via Feshbach resonances would enable a calibrated test of the predicted response of the extracted cubic coefficient \cite{chin2010feshbach}. 
The realization could happen in free fall or in microgravity \cite{becker2018maiusscience,lachmann2021spaceAI}. 
A practical constraint is that the density must be sampled finely enough to resolve the interference fringe, experimentally set by the optical imaging resolution. The main-lobe fringe spacing is of order the Airy scale $x_0$, which at the representative $F=mg$, giving $x_0\approx0.30\,\mu\mathrm{m}$, lies below the ${\sim}1.3\,\mu\mathrm{m}$ resolution of high-resolution quantum-gas imaging~\cite{bennie2013}, but this is a matter of scale rather than principle. The initializing force sets the scale $x_0$ for the free evolution, so a smaller one gives a larger $x_0$ and enlarges the fringe: it grows linearly with $x_0$ while the evolution time grows as $x_0^2$ and $c_3\propto x_0^{-6}$, so the measurable phase $c_3 t^3$ is invariant and both spatial and temporal features are simply stretched. An initializing force with an acceleration of ${\sim}10^{-2}$--$10^{-3}\,g$ gives $x_0$ of at least $1$--$3\,\mu\mathrm{m}$, lifting the fringe above the optical resolution at fixed measurable phase, at the cost of longer free evolution, above tens of milliseconds, naturally accommodated in microgravity, whose residual accelerations reach down to ${\sim}10^{-6}\,g$~\cite{Raudonis2023MicrogravityFacilities}. The non-dispersing, self-healing Airy main lobe is well suited to this long-time, large-scale regime.

Given the $T^3$ scaling of the cubic phase, such interferometric access to the cubic term may also be relevant, through its force-induced component, for precision acceleration measurements, provided that interaction-induced phase contributions are well calibrated \cite{zimmermann2017,amit2019}. 
Furthermore, this experiment would extend interferometric observations of the cubic phase to self-accelerating ultracold quantum fluids. 
In the longer term, the Ai\mbox{-}G collision could serve as a simple interference-based benchmark for weak mean-field nonlinearities via deviations of the extracted cubic coefficient from its non-interacting $g=0$ value. 
More broadly, the force-induced cubic-in-time phase, the linear-potential counterpart of the intrinsic phase carried by Airy eigenstates under free evolution, is tied to the action of a particle in a uniform force field \cite{pellner2026avs,pellner2026thesis}, providing a clean phase observable for comparing accelerated-frame descriptions in matter-wave interferometry, and may therefore inform discussions of equivalence-principle tests in atom interferometry \cite{rosi2017wep}.

\section*{Acknowledgments}
We thank N. Schöneberg and K. D. Follmann for productive discussions. 
G.G.R.\ acknowledges additional support from the MIT School of Science Research Innovation Seed Fund and the C.L.E Moore Instructorship in Applied Mathematics.
Generative AI was used for copyediting and internal consistency checks of the manuscript.

\section*{Contributions}
Manuscript, simulations, and data analysis by M.P. Conceptualization, supervision, manuscript revision and experimental protocol by G.G.R.

\section*{Competing Interests}
The authors declare no competing interests.

\section*{Correspondence}
Georgi Gary Rozenman: \texttt{gary95@mit.edu}

\appendix
\numberwithin{equation}{section}
\renewcommand{\theequation}{\thesection\arabic{equation}}

\titleformat{\section}
  {\centering\normalfont\normalsize\bfseries}
  {Appendix \thesection:}
  {0.6em}
  {}
\titlespacing*{\section}{0pt}{3.0ex}{0.8ex}

\section{Preparation protocol}
\label{app:preparation}

An expanded TEM$_{00}$ mode of waist $w_L$ illuminates a spatial light modulator imprinting the cubic phase $(X/X_s)^3/3$ in the transverse mask coordinate $X$, and a lens of focal length $f$ then performs the optical Fourier transform, so that at its focal plane the field is the finite-energy Airy beam $E_A(x)\propto\mathrm{Ai}(x/x_0)\,e^{a x/x_0}$~\cite{siviloglou2007ol,siviloglou2007prl}, real up to a global phase and a displacement of order $a^2$, with the sign alternating between lobes. The mask and the illumination set all parameters,
\begin{equation}
  x_0=\frac{\lambda f}{2\pi X_s},
  \qquad
  a=\Big(\frac{X_s}{w_L}\Big)^{2},
\end{equation}
with $\lambda$ the optical wavelength, the mask scale $X_s$ fixing the Airy scale, the illuminating waist $w_L$ the truncation, and a linear mask tilt only the position.

The field is written onto the condensate by a weak two-photon Raman pulse coupling two internal states $|g_1\rangle\to|g_2\rangle$, with $E_A$ as one arm and a flat, co-propagating reference beam as the other, so that the differential wavevector has negligible projection along $x$ and no momentum is imparted. To first order the transferred component is
\begin{equation}
\begin{aligned}
  \psi_{\mathrm{out}}(x) &\propto \Omega(x)\,\psi_0(x) \\[0.3ex]
  &\propto \mathrm{Ai}(x/x_0)\,e^{a x/x_0}\,e^{-x^{2}/4\sigma_0^{2}}, \\[0.3ex]
  \Omega &\propto E_A E_{\mathrm{ref}}^{*} ,
\end{aligned}
\end{equation}
which reduces to the truncated Airy state of Eq.~(8) when the parent condensate, a Gaussian ground state of the initial width $\sigma_0$ of Sec.~2.2, is broad compared to the truncated-Airy extent $L_{\mathrm{Ai}}\sim x_0/a$, that is $\sigma_0\gg L_{\mathrm{Ai}}$, the regime assumed here. The coherence of the transfer is essential, as intensity-based, dissipative shaping is blind to the lobe signs, equivalently to the cubic spectral phase, without which the packet does not self-accelerate. Amplitude and phase transfer of structured light onto condensate components by stimulated Raman coupling is established~\cite{andersen2006,wright2008}, optical phase imprinting on condensates is standard~\cite{dobrek1999,denschlag2000}, and related Airy-preparation schemes are proposed in Ref.~\cite{efremidis2013}.

Because the transferred packet occupies $|g_2\rangle$, the bare total density contains no cross term with the parent. A spatially uniform $\pi/2$ mixing pulse applied before imaging projects both components onto a common basis and restores the interference, its phase directly implementing the controlled offsets $\alpha$ required by the phase-stepped reconstruction of Sec.~2.5. The untransferred parent thus serves as the mutually coherent Gaussian reference of the Ai-G geometry, with a relative phase reproducible between repetitions. For the Ai-Ai collision, two sign-mirrored masks produce
\begin{equation}
\begin{aligned}
  \psi_{1}(x) &\propto \mathrm{Ai}\big((x-\ell_1)/x_{0,1}\big)\,e^{a_1 (x-\ell_1)/x_{0,1}}, \\[0.3ex]
  \psi_{2}(x) &\propto \mathrm{Ai}\big(-(x-\ell_2)/x_{0,2}\big)\,e^{-a_2 (x-\ell_2)/x_{0,2}} ,
\end{aligned}
\end{equation}
both in $|g_2\rangle$, which interfere directly and self-accelerate toward one another. The two initial centres are placed symmetrically, $\ell_1=-\ell_2$, the parent can afterwards be removed by a resonant push, and the scale asymmetry $x_{0,2}=1.3\,x_{0,1}$ of Fig.~\ref{fig:airy_collision} enters through the masks alone. The beams vary only along $x$, consistent with the quasi-one-dimensional reduction of Sec.~2. The projection optics limit the structuring to $\delta x\simeq\lambda/(2\,\mathrm{NA})$, with NA the numerical aperture, so this route favors Airy scales $x_0$ of order a micron or larger.

\section{Uncertainty Estimation}
\label{app:uncertainty}

In order to test the nonlinearity influence experimentally, 
even though the reliable valley structure of the robustness heatmap likely exists,
choosing one best fit window for an accurate estimator $\hat c_3$ is naive despite an accurate calibration of $g=0$.
In general, suboptimal conditions should be assumed to maintain a conservative attitude, and therefore the  experimental circumstances might create noisy heatmaps with enhanced small-scale fluctuations – generally meaning a worse signal-to-noise ratio of the underlying phase data.
Thus, not a single-window estimate should be selected since the cubic coefficient is hypersensitive even to small time interval changes, but rather an average over several values - for instance by simply fixing the value $t_{\mathrm{start}}$ and varying the upper boundary $t_{\mathrm{end}}$, to avoid selecting a noise-induced minimum instead of the perfect minimum in an ideal valley cross-section.
The sampling range should, as far as possible, be centered around the valley structure and should ideally contain only a limited number of values. The choice of this range should be based on a trade-off between deliberately including points with small deviations, which may cancel due to positive and negative deviations around the minimum, and ensuring a sufficient number of data points to cope with noise.

However, naive simple averaging would intrinsically yield a mean with an enormously underestimated uncertainty since these coefficients rely on data windows which are highly correlated, and hence reasonably concluding that $c_3$ is subjected to the same correlations.
At this point, it is worth noting that – as all the heatmaps illustrate well – the hypersensitivity of the cubic coefficient does not compensate the correlation effects, apart from the necessary perfect match of their contrary magnitude of both effects, because it refers to the accuracy and not its precision, the uncertainty. 
In case of a single-point-based estimator, a noise-induced, slightly displaced position in the heatmap not providing the perfect fit window with the true $c_3$ value can heavily distort succeeding nonlinearity measurements through a bias because of the hypersensitivity.
Even neighbouring points in the heatmap can have noticeably other values, again reminding of the compressed representation of the deviations by the colours, which enable the overview of rapid changes.

In order to obtain a realistic uncertainty estimate of the algorithm-based systematics, the heteroskedasticity and autocorrelation consistent (HAC) Newey-West estimator is used \cite{neweywest1987}. It considers correlation by introducing corrections to the covariance matrix. 
Each fit of the relative phase data over a certain time interval yields an
estimate $c_{3,i}$ with an associated fit uncertainty $\mathrm{SE}_i$.
Then, the unweighted mean $\mu = \frac{1}{m}\sum_{i=1}^m c_{3,i}$ is constructed to determine
the corresponding residuals $r_i = c_{3,i} - \mu$.
Thus, the autocovariances up to lag $L$ can be defined as
\begin{equation} 
  \hat\gamma_\ell
  = \frac{1}{m-\ell} \sum_{i=1}^{m-\ell} r_i\, r_{i+\ell},
  \qquad \ell = 0,1,\dots,L ,
\end{equation}
to account for serial correlation between neighbouring windows, and subsequently applying the Bartlett weights
\begin{equation}
  w_\ell = 1 - \frac{\ell}{L+1}, \qquad
  \tilde\gamma_\ell = w_\ell \hat\gamma_\ell \ .
\end{equation}
From $\tilde\gamma_\ell$, a Toeplitz process covariance
matrix is assembled,
\begin{equation}
  (\Sigma_{\mathrm{proc}})_{ij} =
  \begin{cases}
    \tilde\gamma_{|i-j|} & |i-j| \le L, \\
    0                    & \text{otherwise}.
  \end{cases}
\end{equation}
The heterogeneous fit uncertainties enter as an additional diagonal term,
\begin{equation}
  \Sigma = \Sigma_{\mathrm{proc}} +
  \operatorname{diag}\!\bigl(\mathrm{SE}_1^2, \dots, \mathrm{SE}_m^2\bigr),
\end{equation}
so that $\Sigma$ represents a realistic covariance matrix of the vector
$\mathbf{c} = (c_{3,1},\dots,c_{3,m})^{\mathsf T}$, combining process correlations and window-dependent
measurement noise.
Eventually, the standard error is given by
\begin{equation}
  \mathrm{SE}(\hat\mu) =
  \left(\mathbf{1}^{\mathsf T}\Sigma^{-1}\mathbf{1}\right)^{-1/2},
\end{equation}
with the mean estimate $\hat\mu =
  \frac{\mathbf{1}^{\mathsf T}\Sigma^{-1}\mathbf{c}}
       {\mathbf{1}^{\mathsf T}\Sigma^{-1}\mathbf{1}}$ and $\mathbf{1}=(1,\dots,1)^{\mathsf T}$.
This approach yields a conservative and data-driven uncertainty estimate, which considers both serial correlation and heterogeneous fit uncertainties, and which should be regarded only as a lower bound of the actual experimental uncertainty – focusing here only on the algorithm-systematics.

Generally, since the majority of underlying data are shared with the adjacent window, substantial serial correlation is expected, motivating the use of relatively large lags $L$, up to $L=m-1$.
However, for $L$ approaching $m$, the number of terms contributing to each autocovariance estimate in Eq.~(B1) decreases, making high-lag estimates of the autocovariances increasingly noisy. Thus, finite window sizes of the averaging interval can limit accounting properly for strong correlation.
The Bartlett weights downweight these high-lag contributions, which helps to regularize the HAC covariance estimate and to keep it well behaved – in particular, the Newey--West construction yields a positive semidefinite estimate \cite{neweywest1987}. 
In practice, accounting for long-range correlations can also lead to partial cancellation in the residual products summation because of opposing signs, causing negative contributions, so that increasing $L$ does not necessarily increase the estimated uncertainty, here due to the heatmap valley structure.  
Because $c_3$ is highly sensitive to the selected interval, the mapping from window overlap to correlations in $c_3$ need not be straightforward. 
The HAC Newey–West estimate should therefore be interpreted empirically as a conservative, data-driven standard-error correction for window-induced correlation, rather than a complete account of all algorithmic, protocol-induced uncertainties.

For a careful analysis, the $L$, that maximizes the uncertainty is chosen,
which is exemplified with
$\hat c_{3,\mathrm{HAC}} = (1.160 \pm 0.025)\times 10^{11}\,\mathrm{rad/s^3}$
for $L=14,\ m=28$ in Table~\ref{tab:hac-sensitivity}.
Hence, the Newey-West-corrected standard error is increased by  $31.58 \, \% $.
The theoretical value $c_{3,\mathrm{theo}}$ lies within the corresponding         
one-sigma interval and thus agrees with this Newey–West estimate.

\begin{table}[h]
  \centering
  \caption{Heteroskedasticity and autocorrelation consistent uncertainty estimate for a fixed start time
  $t_{\mathrm{start}}=0.015~\mathrm{ms}$ with $m=28$ windows, for the
  Ai-G collision with DPE.}    
  \begin{tabular}{rrr}
    \hline\hline
    $L$ & $ \hat c_{3,\mathrm{HAC}}\,[10^{11}\,\mathrm{rad/s^3}]$ & $\mathrm{SE}\,[10^{9}\,\mathrm{rad/s^3}]$ \\
    \hline
     3 & 1.16113 & 1.768 \\
     6 & 1.16081 & 2.146 \\
     8 & 1.16066 & 2.302 \\
    11 & 1.16053 & 2.442 \\
    14 & 1.16046 & 2.494 \\
    17 & 1.16039 & 2.468 \\
    20 & 1.16034 & 2.387 \\
    22 & 1.16033 & 2.307 \\
    25 & 1.16032 & 2.140 \\
    \hline\hline
  \end{tabular}
  \label{tab:hac-sensitivity}
\end{table}

\section{Airy-scale parameter sweep}
\label{app:airysweep}

\begin{figure}[H]
  \centering
  \includegraphics[width=\columnwidth,
    trim=15 334 75 100,clip]{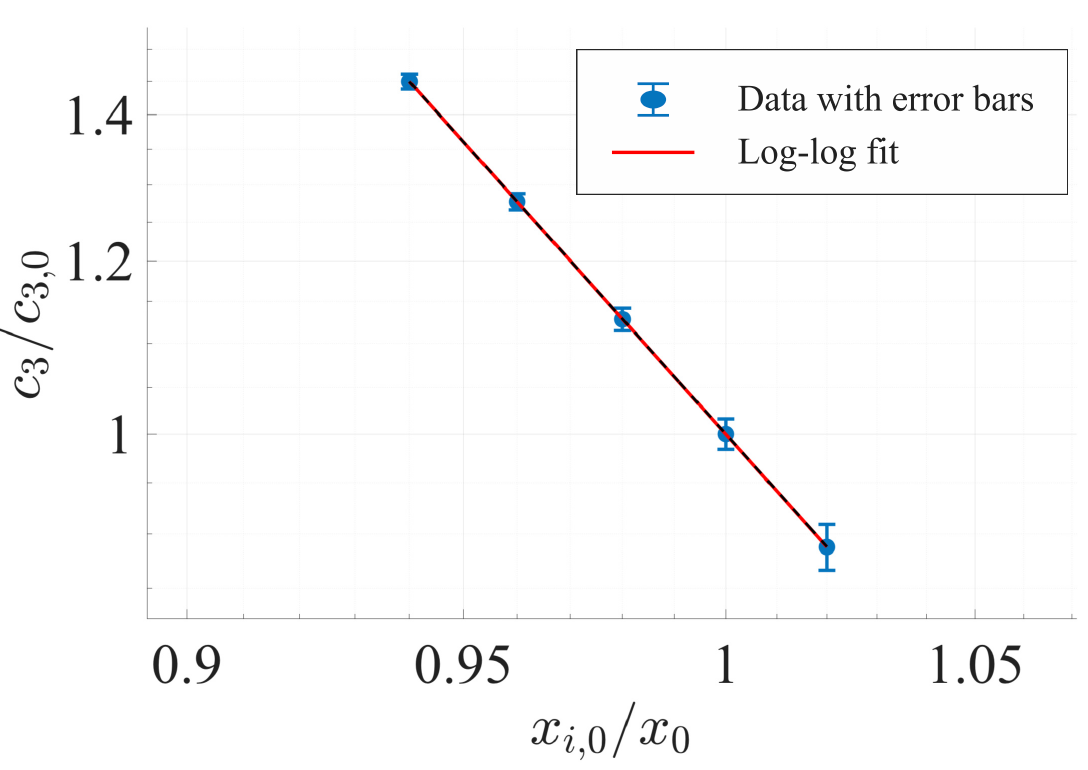}
  \caption{
  Log-normalized cubic coefficient $\log\!\bigl(c_3/c_{3,0}\bigr)$ for the Ai-G scenario (HPE) at $g=0$, shown as a function of the Airy scale $x_{i,0}$.}
  \label{fig:parameter_scan}
\end{figure}

As an independent validation of the $c_3$ predictions obtained from the robustness heatmaps, 
the slope of $\log\!\bigl(c_3/c_{3,0}\bigr)$ in Fig.~\ref{fig:parameter_scan} is found to be $-6.005 \pm 0.004$, 
confirming the predicted scaling $c_3 \propto x_0^{-6}$ in Eq.~21.

\begingroup
\footnotesize                 
\setlength{\bibsep}{0pt}      
\setlength{\itemsep}{0pt}     
\setlength{\parskip}{0pt}
\setlength{\bibhang}{1.2em}   

\bibliographystyle{unsrt}     
\bibliography{references}
\endgroup

\end{document}